\documentclass[10.5pt,compsoc]{BD}
\usepackage{graphicx}
\usepackage{footmisc}
\usepackage{subfigure}
\usepackage{url}
\usepackage{multirow}
\usepackage[noadjust]{cite}
\usepackage{amsmath,amsthm}
\usepackage{amssymb,amsfonts}
\usepackage{booktabs}
\usepackage{color}
\usepackage{ccaption}
\usepackage{booktabs}
\usepackage{float}
\usepackage{fancyhdr}
\usepackage{caption}
\usepackage{xcolor,stfloats}
\usepackage{comment}
\graphicspath{{figures/}}
\usepackage{cuted}  
\usepackage{captionhack}
\usepackage{epstopdf}
\usepackage[caption = true]{subfig}

\headevenname{\zihao{-5}{\textbf{\emph{}}} 2021}%

\headoddname{{\sf Thosini Bamunu Mudiyanselage et al.:}\quad {\textbf{\emph{Graph Convolutional Neural Network based Covid Detection using CXR}}}}%

\newtheoremstyle{mystyle}{0pt}{0pt}{\normalfont}{1em}{\bf}{}{1em}{}
\theoremstyle{mystyle}
\renewcommand\figurename{Fig.~}

\newcommand{\nop}[1]{}

\makeatletter
\renewcommand{\@biblabel}[1]{[#1]\hfill}
\makeatother

\begin{document}

\hyphenpenalty=50000

\makeatletter
\newcommand\mysmall{\@setfontsize\mysmall{7}{9.5}}

\newenvironment{tablehere}
  {\def\@captype{table}}
  {}
\newenvironment{figurehere}
  {\def\@captype{figure}}
  {}

\thispagestyle{plain}%
\thispagestyle{empty}%

\let\temp\footnote
\renewcommand \footnote[1]{\temp{\zihao{-5}#1}}
{}

\vskip .2mm\noindent
{\zihao{5-}\textbf{\scalebox{1}[1.0]{\makebox[5.6cm][s]{%
\color{white}{V\hfill o\hfill l\hfill u\hfill m\hfill%
e\hspace{0.356em}1,\hspace{0.356em}N\hfill u\hfill%
m\hfill b\hfill e\hfill r\hspace{0.356em}1,\hspace{0.356em}%
S\hfill e\hfill p\hfill t\hfill e\hfill%
m\hfill b\hfill e\hfil lr\hspace{0.356em}2\hfill0\hfill1\hfill8}}}}}\\

\begin{strip}
{\center
{\zihao{3}\textbf{
Covid-19 Detection from Chest X-ray and Patient Metadata using Graph Convolutional Neural Networks}}
\vskip 9mm}

{\center {\sf \zihao{5}
Thosini Bamunu Mudiyanselage, Nipuna Senanayake, Chunyan Ji, Yi Pan and Yanqing Zhang
}
\vskip 5mm}
%

\centering{
\begin{tabular}{p{160mm}}

{\zihao{-5}
\linespread{1.6667} %
\noindent
\bf{Abstract:} {\sf
The novel corona virus (Covid-19) has introduced significant challenges due to its rapid spreading nature through respiratory transmission. As a result, there is a huge demand for Artificial Intelligence (AI) based quick disease diagnosis methods as an alternative to high demand tests such as Polymerase Chain Reaction (PCR). Chest X-ray (CXR) Image analysis is such cost-effective radiography technique due to resource availability and quick screening. But, a sufficient and systematic data collection that is required by complex deep leaning (DL) models is more difficult and hence there are recent efforts that utilize transfer learning to address this issue. Still these transfer learnt models suffer from lack of generalization and increased bias to the training dataset resulting poor performance for unseen data. Limited correlation of the transferred features from the pre-trained model to a specific medical imaging domain like X-ray and overfitting on fewer data can be  reasons for this circumstance. In this work, we propose a novel Graph Convolution Neural Network (GCN) that is capable of identifying bio-markers of Covid-19 pneumonia from CXR images and meta information about patients. The proposed method exploits important relational knowledge between data instances and their features using graph representation and applies convolution to learn the graph data which is not possible with conventional convolution on Euclidean domain. The results of extensive experiments of proposed model on binary (Covid vs normal) and three class (Covid, normal, other pneumonia) classification problems outperform different benchmark transfer learnt models, hence overcoming the aforementioned drawbacks.

 }
}

\end{tabular}
}
\vskip 6mm

\vskip -3mm
\zihao{6}\end{strip}

\thispagestyle{plain}%
\thispagestyle{empty}%
\makeatother
\pagestyle{tstheadings}

\begin{figure}[b]
\vskip -6mm
\begin{tabular}{p{44mm}}
\toprule\\
\end{tabular}
\vskip -4.5mm
\noindent
\setlength{\tabcolsep}{1pt}
\begin{tabular}{p{1.5mm}p{79.5mm}}

$\bullet$& Thosini Bamunu Mudiyanselage, Nipuna Senanayake, Chunyan Ji, Yi Pan and Yanqing Zhang are with the Department of Computer Science, Georgia State University, Atlanta, GA 30302, USA. 

E-mail: tbamunumudiyanselag1@student.gsu.edu; 
\end{tabular}
\end{figure}\zihao{5}

\vbox{}
\vskip 1mm
\zihao{5}

\section{Introduction}
\label{s:introduction}
\noindent
Covid-19 is a virus from Corona virus family and scientifically named as Severe acute respiratory syndrome coronavirus 2 (SARS-CoV-2). Respiratory transmission from person to person has caused  rapid spreading of the virus and has lead to a global pandemic situation. Infection of Corona viruses causes pneumonia and in more sever cases can cause multi-organ failures and even death. This is a major health problem as Covid-19 has been first identified in humans in 2019 while it had already started spreading at the time of discovery. Reverse transcription polymerase chain reaction (RT-PCR) is the test frequently used for diagnosing the patients. But due to the exponential increase of demand for tests, the availability of test kits is insufficient. Additionally these tests are time consuming and a manual process with high false positives. This results in a situation where actual patients cannot be quickly identified and isolated in order to minimize the spread and infecting others. Failure to do so leads to a collapse in health systems of many countries due to exceeding the maximums of their capacities. 
As a solution, there is a high demand for automated quick screening methods such as the use of chest radiography imaging. Resource availability, cost effectiveness, quick screening time are major advantages of these automated radiography screening methods. Thus, quick diagnosis of those who are showing Covid-19 pneumonia patterns in CXR can be isolated and treated quickly to mitigate the risk of infecting a larger population. Moreover, these automated image analysis tools can be used to detect and monitor progression of lung damages of Covid-19 patients. 

A collection of a larger datasets of these CXR images is a requirement for better performance in deep learning but hard to achieve given the lack of available data and data is being collected on the fly. Also, labeling these images manually is a laborious task. As a result, transfer learning has been introduced and currently being used in many medical image processing tasks such as brain images, retina images and X-ray images analysis. In transfer learning, deep neural networks are trained on a large benchmark datasets \cite{1} and then fine tuned on a specific task with fewer labels than the pre-training problem. A study \cite{2} suggests transfer learnt CNN architectures to detect common pneumonia, Covid-19, and normal incidents on a small X-Ray image dataset. Another work in \cite{3} developed an automated deep transfer learning-based approach for detection of COVID-19 infection in chest X-rays by using the extreme version of the Inception (Xception) model. Five pre-trained convolutional neural network based models (ResNet50, ResNet101, ResNet152, InceptionV3 and Inception-ResNetV2) have been proposed in \cite{4} for the detection of coronavirus pneumonia infected patients using chest X-ray radiographs and above models were evaluated on three different binary classifications with four classes (COVID-19, normal (healthy), viral
pneumonia and bacterial pneumonia) using 5-fold cross validation. Due to the small COVID-19 dataset availability, another study was carried out to classify chest computerized tomography (CT) scans as covid infected or not using DenseNet201 based deep transfer learnt automated tool in \cite{5}. The \cite{6} is also utilized improved AlexNet based model to classify CT scans as Covid-19 and healthy samples. CT scans examine inner soft tissues of organs as well but x-ray machines use low ionizing radiations and hence X-rays machines are faster, cheaper and less harmful solutions than the CT scan machines. As a result, there are many efforts made on X-ray image datasets to build transfer learnt automated tools to identify covid related signatures appeared in CXR images. On the other hand, this low radiation quality of CXR makes analyzing these covid patterns is difficult and a more error prone task. This further implies the importance of strong automated model's support in diagnosing CXR as covid infected or not.

Another work in \cite{7} proposed three pre-trained CNNs, AlexNet, GoogleNet, and SqueezeNet which were fine-tuned without data augmentation to carry out 2-class and 3-class classification has rapid training and can contribute to the urgent need for harnessing the pandemic. Meanwhile \cite{8} and \cite{9} suggest the use of pre-trained VGG-16 models along with conventional and data augmentation methods to classify CXR into covid-19, healthy and other pneumonia types and attention based VGG-16 for 3 and 4 class classifications respectively. With suitable selection of training parameters, these models have given good results compared to other complex Deep learning models. The \cite{10} suggests a patch-based convolutional neural network approach on segmented small number of CXR instances as a solution for less availability of well-curated training data samples. This study adopted pre-trained ResNet-18 model for patch training. Only the \cite{11} has combined CXR images from five different sources to create a bigger data set of about 14000 CXRs instead of transfer learning, but reported performance is not much higher due to various artifacts related to different devices that produce images and inbuilt noises. Moreover, \cite{10} above with a small dataset showed higher prediction capabilities compared to \cite{11}. 

However, most of transfer learnt models fail to learn a generalized representation of data or tend to show a bias towards training dataset as these models are trained on a small number of instances. Hence, these models end up having poor performance on unseen data. In this study, we propose a graph convolution neural network to detect Covid-19 patients using the characteristics appeared on CXR images. The purpose of this approach is to combine the strength of graph representation to capture the important relational knowledge between different data instances and apply convolution operations to learn the graph structure and eventually classify data instances accurately to the most appropriate class. We observe that deep learning efficiently captures hidden patterns of data and makes better predictions in many domains \cite{12}. Deep learning operations such as convolution which takes simple correlations between pixels within Euclidean data has contributed to the better performance in these data. But graphs are extensively employed data structures which capture interactions between different data instances and have the potential for better learning \cite{13, 14, 15, 16}. As a result, there are recent efforts to extend deep learning operations to graph data which have irregularities such as variable sized and unordered nodes. Hence, new generalizations of important operations from deep learning have been rapidly developed into graphs and various graph Neural Networks and learning algorithms have emerged recently \cite{17}. GCN \cite{18, 19} is a method which performs graph convolutions by aggregating the neighbour nodes' information similar to 2D convolution.

Thus, the contribution of novel GCN based Covid-19 detection model on CXR can be summarized as below: 

\begin{itemize}
\setlength{\itemsep}{1pt}
  \item First, we use a Convolution Neural Network (CNN) encoder to extract features for all the samples in the CXR dataset. 
  \item Next, compute the similarity matrix of data instances using extracted features and build the graph fusing similarity matrix and meta-data information. 
  \item A GCN is then adopted on the built graph to learn the graph structure and high level features to classify data samples into appropriate classes.  
  \item Moreover, the propose method diagnoses Covid-19 and differentiate Covid-19 pneumonia patterns from other pneumonia types including bacterial, tuberculosis, fungal and other viral types.  
\end{itemize}

Hence, the proposed model overcomes above mentioned limitations and the results of extensive experiments demonstrate that GCN based Covid-19 detection on CXR outperforms ubiquitous transfer learning approaches on CXR in disease diagnosis. Also, most of the existing works are limited for binary classification \cite{4, 5, 6} while there is an appealing need for models to discriminate Covid-19 from other lung infections. In our study, the proposed GCN based model is adopted on binary and three class classification where the robustness of the model is further evaluated on five classes as well. 

The rest of this paper is organized as follows: Section 2 describes details of CXR images and datasets, generalized feature representation of transfer learnt models and CNN encoder which is used to encode images, section 3 expresses  different steps of the proposed method including building the graph using encoded feature similarity plus meta-data fusion and applying GCN algorithm. Results of the experiments are presented in Section 4. Finally in Section 5, the Summarization and Conclusion are included. 

\section{Data and CNN Encoder for Feature Extraction}
\label{s:CNN Encoder for Feature Extraction}
\noindent

Here we study the feature representations learnt by different transfer learnt models and other CNN models. One purpose of this study is to learn the affect of  transfer learning for feature representations compared to a CNN model with random initialization. It also helps to identify a CNN encoder which gives generalized feature representations for different CXR images in order to build the graph based on feature similarities of data instances. This study is preceded by a description of CXR images and the datasets we adopted in this work.

\subsection{Data}
\noindent

The CXR images have been retrieved from publicly available two repositories \cite{20} and \cite{21}. The data at \cite{20} contains CXR images of patients with Covid-19, other viral infections like Middle East Respiratory Syndrome (MERS), Sever Acute Respiratory Syndrome (SARS). Further, it includes cases of bacterial infections like Tuberculosis and fungal infected patients' CXR images. In addition to the CXR images, meta data which consists of patients' information such as offset (the length of the stay from hospitalization to the time of test in days),age, gender, survival, RT-PCR,  whether they were in ICU, currently in ICU, currently intubating, intubated and whether supplemental $O_2$ is needed were also utilized in our GCN model. All together, there are 10 features extracted from meta data. As there are few normal CXR images in \cite{20}, we obtained normal CXR images from Kaggle repository \cite{21}. 

Two datasets DS1 and DS2 were constructed using extracted data from the above two data repositories for training and testing proposed model and other pre-trained models. DS1 includes about 150 CXR of Covid-19 and 150 CXR of normal subjects which was designed for binary classification problem. DS2 was designed for three class classification where we differentiate Covid-19 infected subjects from normal and other pneumonia type infections. This dataset consists of 150 Covid-19, 150 other pneumonia and another 150 instances for normal CXR images. Within the samples of other pneumonia CXR, there are cases of bacterial infections, fungal infections and other viral infections of subjects. Figure 1 presents representative CXR images for Covid-19, bacterial, fungal, other viral and normal patients respectively.  

\begin{figure*}[ht]
  \subfigure[Covid-19]{
  \begin{minipage}{.19\textwidth}
  \centering
  \includegraphics[width=3cm, height=3cm]{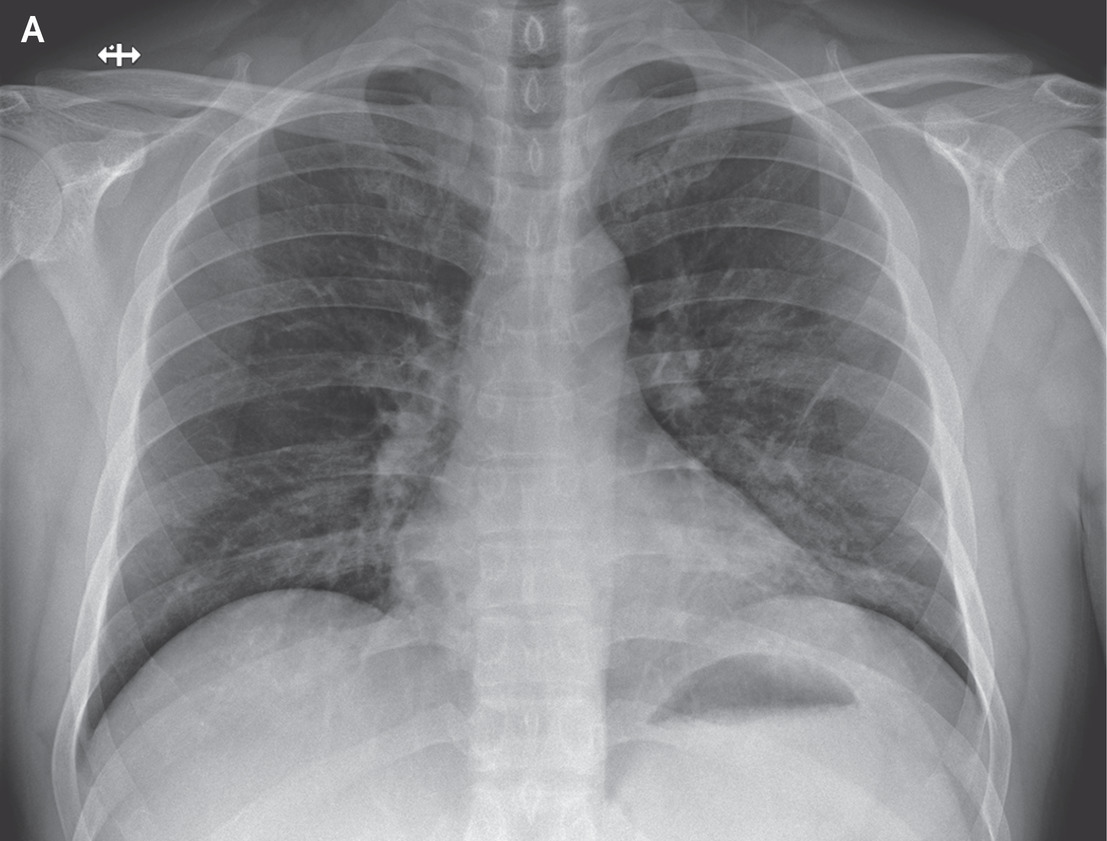}
  \end{minipage}\hfill}
  \subfigure[Bacterial]{
  \begin{minipage}{.19\textwidth}
  \centering
  \includegraphics[width=3cm, height=3cm]{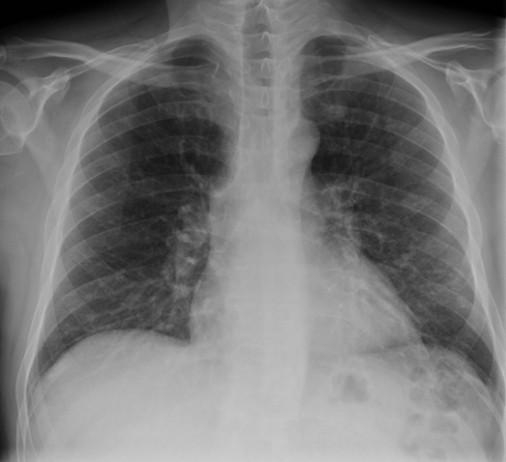}
  \end{minipage}\hfill}
  \subfigure[Fungal]{
  \begin{minipage}{.19\textwidth}
  \centering
  \includegraphics[width=3cm, height=3cm]{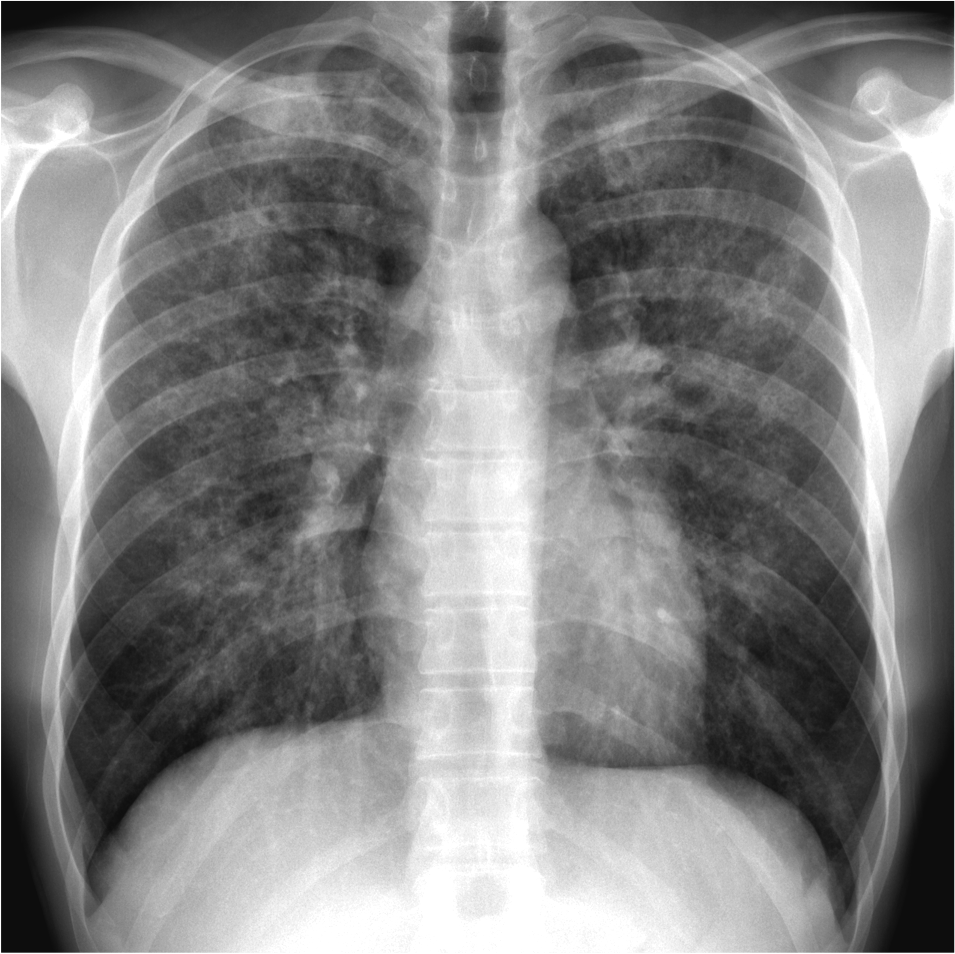}
   \end{minipage}\hfill}
   \subfigure[Viral]{
  \begin{minipage}{.19\textwidth}
  \centering
  \includegraphics[width=3cm, height=3cm]{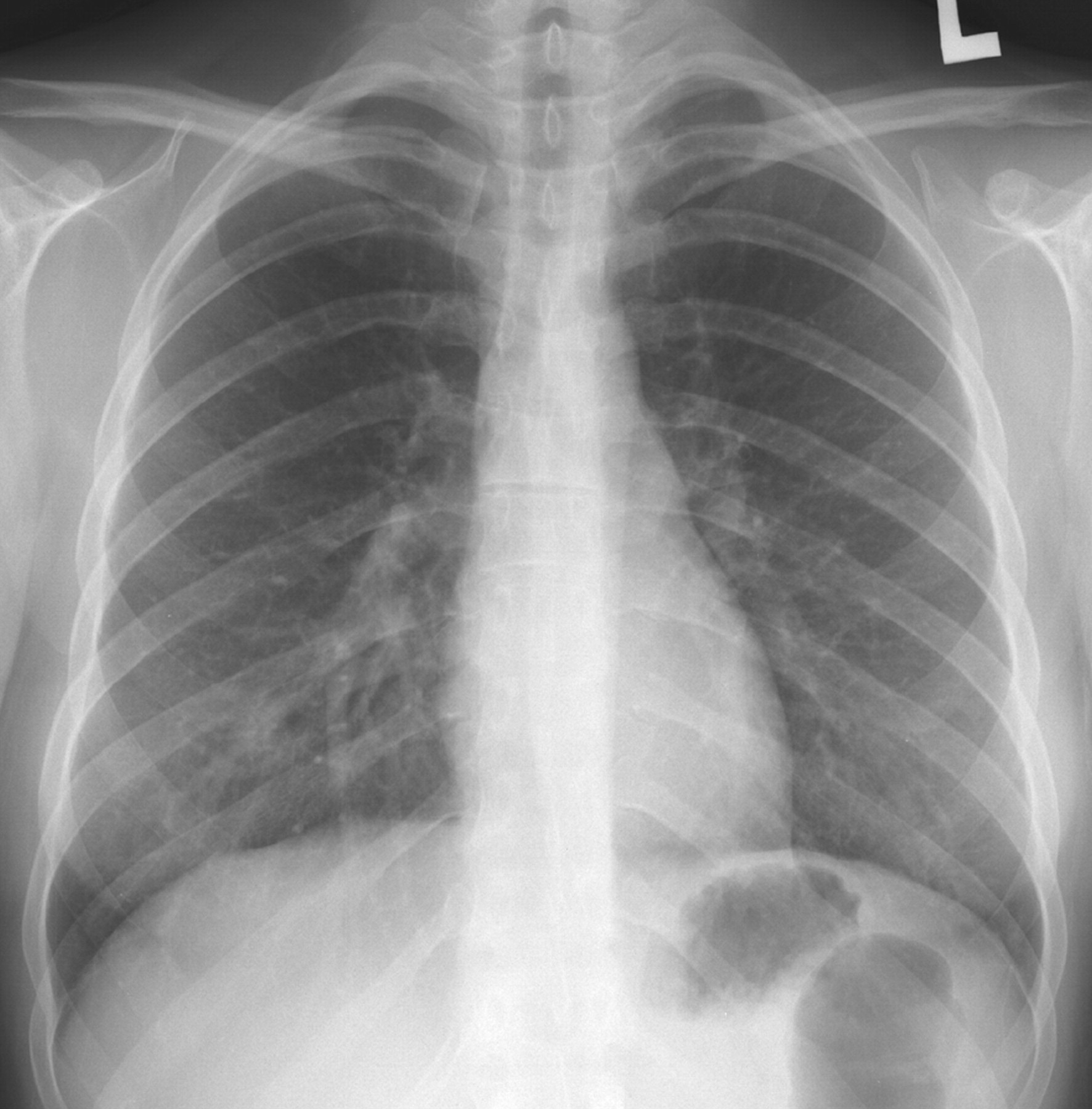}
   \end{minipage}\hfill}
   \subfigure[Normal]{
  \begin{minipage}{.19\textwidth}
  \centering
  \includegraphics[width=3cm, height=3cm]{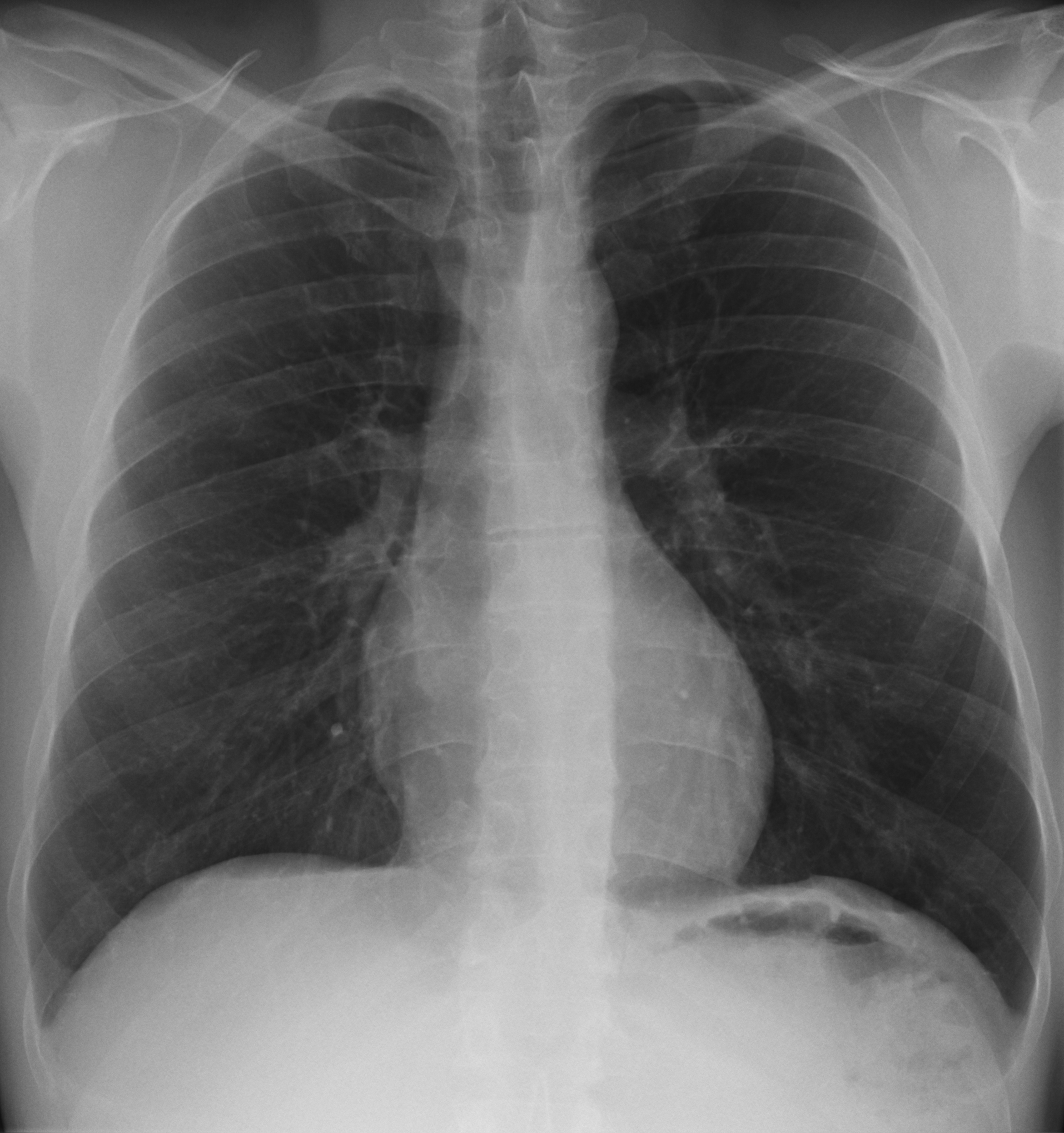}
   \end{minipage}\hfill}
  \caption{Representative CXR images from datasets corresponding to different infections}
\end{figure*}

In these sample CXR images, there are certain areas with hazy opacification or increased attenuating as these areas are filled with some substances other than air. This is an indication of displacement of air by fluid and a collapse of periphery of lungs due to various infections. Though these areas are visible as more grey and cloudy compared to darker areas with lower attenuation, differentiating the cause of infection is not possible for human eye alone. Further, the specific radiographic patterns can be variable depending on different viral strains making it hard even for expert to diagnose using CXR alone. This further implies the demand for efficient and accurate automated CXR based techniques for disease diagnosis. 

\subsection{Feature representation for CXR}
\noindent

As we stated earlier, there is a frequent use of transfer learnt models for medical image processing tasks including CXRs. It is worth studying the affect of transfer learning on feature representation learning. Therefore, we used three benchmark transfer learnt models VGG16, InceptionV3 and ResNet-50 to compare the hidden representations learnt by these models for the same dataset. As a baseline, we used another CNN with three convolution layers (3-CNN) which starts  with random initialization and it is depicted in Figure 2.  

\begin{figure}[h]
  \centering
  \includegraphics[width=0.45\textwidth]{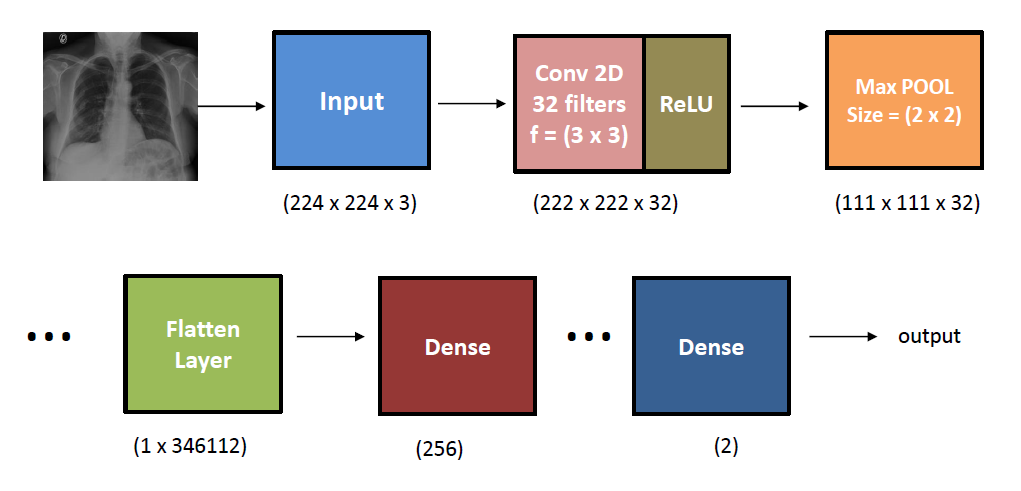}
  \caption{Three layer CNN architecture}
\end{figure}

We calculate the similarity scores of feature representations learnt by above transfer learnt models and the 3-CNN model in multiple training. For that, we extracted the hidden representations from the top layers (before the output) of these models and computed the cosine similarity in between same data instances. Calculated similarities in different configurations of networks trained from Image-Net weights and random weights are plotted in Figure 3.

\begin{figure}[h]
  \centering
  \includegraphics[width=0.45\textwidth]{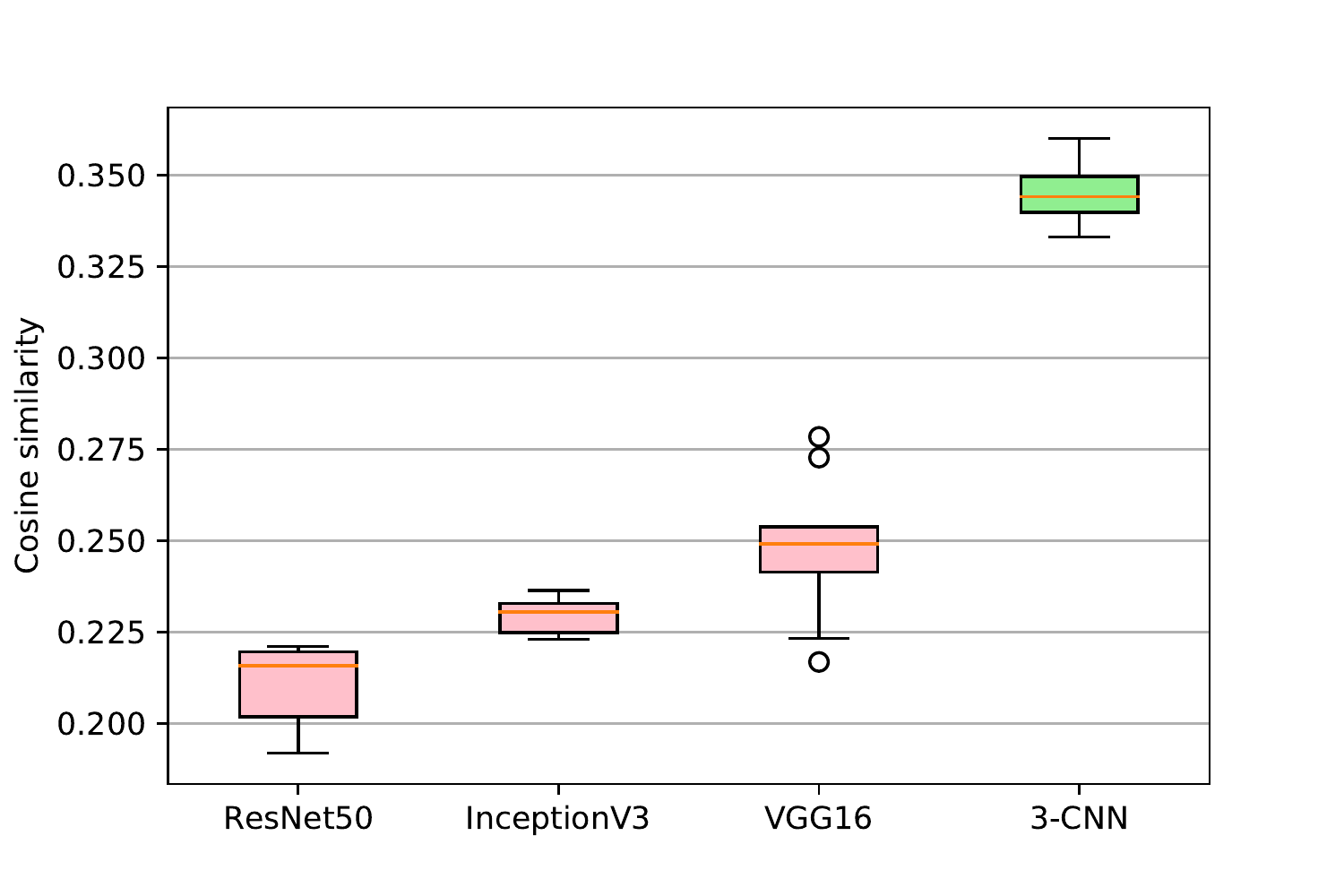}
  \caption{Similarity scores of feature representation between models trained from pre-trained ImageNet weights (pink) and random weight initialization (green)}
\end{figure}

We can observe the feature representations learnt by three benchmark models VGG16, InceptionV3 and ResNet-50 are less similar for the same dataset. For 3-CNN model, there is greater overlap between learnt feature representations, hence leads to comparatively high similarity values. More specifically, 3-CNN has achieved a 10\% similarity improvement which indicates a stable feature learning. Therefore, we use this 3-CNN model (Figure 2) as our encoder to convert the CXR into feature vectors which then we use for graph building.

\section{Proposed GCN based Methodology}
\noindent

This section describes the proposed method with the detailed information regarding construction of graph and adopting GCN based learning for Covid-19 and other pneumonia types detection. The process can be divided into three steps: Encode CXR images into feature vectors, Construct a graph fusing feature vectors and meta data information of patients and Applying graph convolutions to extract high level features of the graph to ultimately classify nodes of the graph into different classes. Overview of the method is depicted in Figure 4.  

\begin{figure*}[h]
  \centering
  \includegraphics[width=\textwidth]{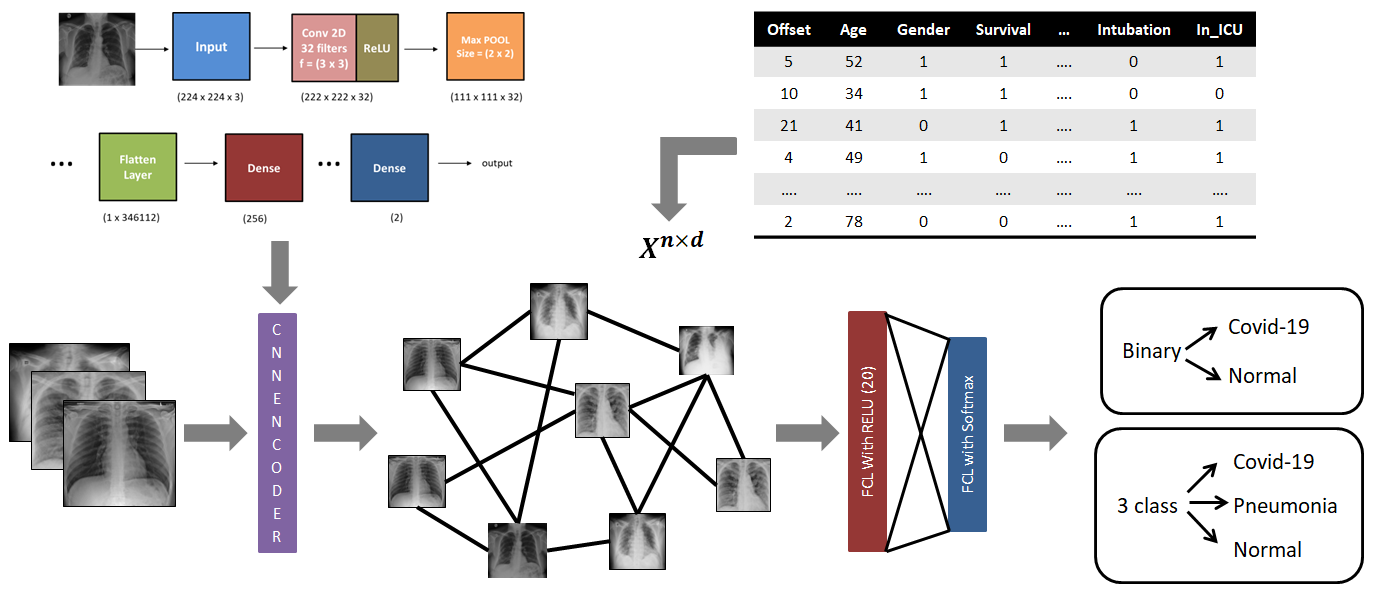}
  \caption{Overview of the proposed Method: 1. Encoding CXR images into feature vectors, 2. Construct a graph fusing feature vectors and meta data information of patients and 3. Applying graph convolutions to extract high level features of the graph to classify nodes into different classes}
\end{figure*}

\subsection{Encoding CXR and building graphs}
\noindent

At first, we convert CXR images into feature vectors as discussed in Section 2.2. Our  3-CNN model is adopted as the encoder to have a generalized feature representation for CXR images. The encoded feature vectors are used to calculate the similarity matrix $S^{ n \times n}$ between all $n$ images. Next, a graph $G=(V,E)$ is built using the calculated feature similarity matrix $S$ where $V=v_{1},...,v_{n}$ which has $n$ number of nodes to represent $n$ CXR images and $E$ contains $m$ number of edges among nodes. We used a threshold $\alpha$ for deciding the connectivity of two nodes. Only if the corresponding similarity value between two given nodes are greater than $\alpha$, these two nodes will be connected. The affect of this threshold on graph connectivity and final classification performance is also discussed in Section 4.3 under Experiments. The structural information of this final graph can be denoted as $A$ and the neighbour nodes of a particular node $v$ is denoted as $N(v)$.  Next we assign the initial features $X^{n\times d}$ for the graph $G$ where $X$ is the metadata information of patients. This includes: age, gender, survival, RT-PCR, currently in ICU, was in ICU, intubation present, intubated, etc. Here we performed data imputation for missing value prediction as well. Our purpose is to map nodes to a lower dimensional feature matrix $Z^{n\times a}$ starting from initial feature matrix $X^{n\times d}$ where $d$ is the starting dimension of each node $n$ and $a$ is less than $d$. Finally, we use these learnt high level features to classify each node $n$ in the graph $G$. For that we adopt GCN that we discuss in the next sub section. 

\subsection{GCN based learning}
\noindent

GCN algorithm can mainly be divided into two steps: Aggregate and Update. The first step which collects the neighbour information is called the aggregation and based on aggregated information of neighbours, updating the current node is the second step which is called update. As this is an iterative process, there should be a feature initialization at the beginning and can be defined as below for node $v$ where $0$ indicates the initial step. 

\begin{equation}
h_v^0 = x_v
\end{equation}

Next step is the aggregation of neighbours' information in order to update the current representation of each node. Equation 2 represents aggregation of embedding vectors of each neighbour node $u$ of the current node $v$.

\begin{equation}
r_v=f_{aggregate}(\{h_u | u \in N (v)\})
\end{equation}

After obtaining $r_v$ (neighbour node's representation), the next step is to update the current node $v$ representation as below. 

\begin{equation}
h_{v}^{k}=f(W_k(r_v,h_{v}^{k-1}))
\end{equation}

Let $h_{v}^{k}$ be the output of $k^{th}$ convolutional layer and a neural network function is used to map previous layer embedding $h_{v}^{k-1}$ to the reduced high level embedding $h_{v}^{k}$. Therefore, $W_k$ is the learnable parameters of the $k^{th}$ layer and $f$ is the activation function such as ReLU. 

We can replace $f_{aggregate}$ with different aggregation functions such as sum, mean and max-pooling where it takes the summation of neighbour nodes' embeddings, average of neighbour nodes' embeddings and maximum embedding out of all neighbour nodes' embeddings respectively. As an example, if $f_{aggregate}$ is replaced by mean function, Equation 2 and Equation 3 can be combined and represented as below:

\begin{equation}
h_{v}^{k}=f\bigg(W_k\bigg(\bigg\{\frac{\sum_{u\in N(v), A}h_{u}^{k-1}}{|N(v)|}\bigg\}, h_{v}^{k-1}\bigg)\bigg)
\end{equation}

A performance comparison between different aggregation functions is also carried out and is discussed in upcoming section. 

In above equations, $k$ means the number of convolutions which indicates how many neighbours to use to compute the node representations in the algorithm. At $k=0$, all the nodes' embeddings are equal to initial feature vectors and eventually learn node embeddings which are multiple hops away from each node. An example where $k$ is set to 2, is given below.

If we consider node $A$ as our targeting node, all nodes are assigned to initial feature vectors at $k=0$. Therefore, $h_A^0$ is the initial feature vector for node $A$ which is equal to $x_A$. Similarly $x_B, x_C, x_D, x_E, x_F$ are assigned to nodes $B,C,D,E$ and $F$ respectively. At $k=1$ layer $h_A^0$, $h_C^0$, $h_E^0$ and $h_F^0$ are aggregated and $h_B^1$ is updated based on the above aggregation. We repeat the aggregate and update functions for the node $C$. In order to get the node embeddings for targeted node $A$ at $k=2$, we aggregate immediate neighbours' node embeddings $h_B^1$ and $h_C^1$ and update $h_A^2$. Thus, each node in above toy example will eventually learn neighbours' and neighbours of neighbours' embeddings. Hence, given $k=2$, graph learns the neighbourhood of two hops away and we can experiment with multiple neighbourhoods for different values of $k$. 

\begin{figure}[h]
  \centering
  \includegraphics[width=0.45\textwidth]{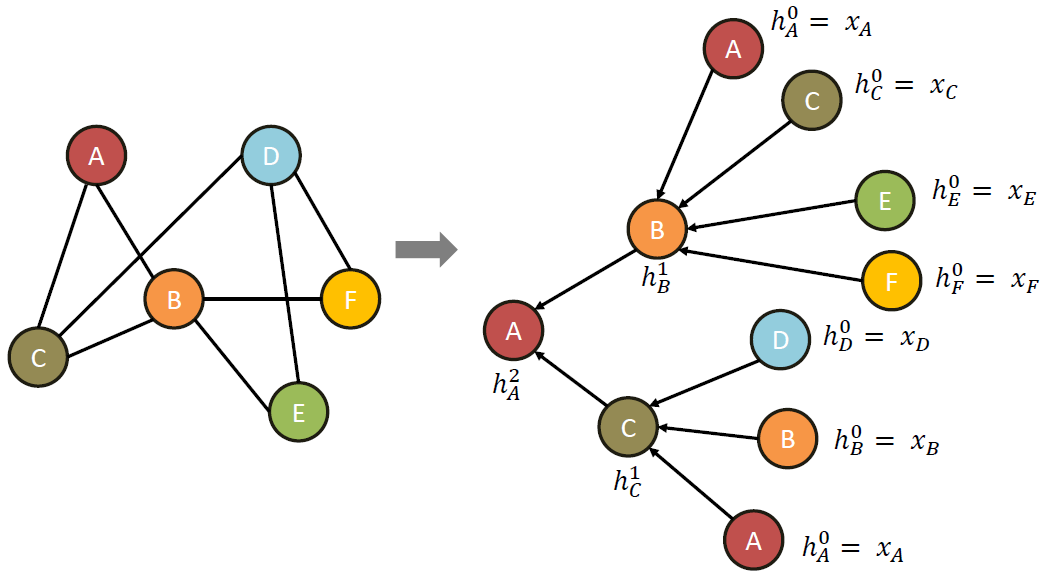}
  \caption{Graph Convolutional learning of two hop neighbourhood for targeted node $A$ }
\end{figure}

However, too many neighbourhoods can reduce the strength of node representation due to the possibility of adding unintended fuzziness and noise to the system. On the other hand, too few neighbourhoods can decrease the non-linearity of the problem making it not suitable for graph learning. The last step of GCN algorithm is to forward the learnt high level feature vector of each node through a softmax layer which finally predicts the probabilities for different classes in the classification. Here, we use training data to build the training graph where we preform the loss calculations based on the output labels and use back propagation to learn the model parameters which we use on test data.         

\section{Experiments and Results}
\noindent

We did various experiments on aforementioned datasets DS1 and DS2 to evaluate the performance of graph convolution algorithm on a binary class classification and three class classification problems for CXR images. To have a fair comparison, we create two graphs called "training graph" and "testing graph" using training testing partition that we performed for the benchmark transfer learnt models. We repeat this process five times and get the average results as the five fold cross validation. At first, we compare the  training and testing performance of GCN based classification on two class and three class problems using training loss and testing accuracies respectively. Next, the performance of GCN is compared with transfer learnt model performance in terms of training loss and testing accuracies on two class and three class classification respectively. Here we present the average results of five fold cross validation on individual classes and also across all classes related to different models. Another test is performed to analyze the results of different aggregation functions of GCN against a range of threshold $\alpha$ values and it is presented in the next section. To evaluate the strength of the proposed GCN, another classification with five classes: covid, bacterial, fungal, other viral and normal was carried out and the final outcome is compared to a transfer learnt ResNet-50 based on confusion matrices. Finally the saliency maps were drawn to demonstrate the effectiveness of important feature detection of transfer learnt models which starts from pre-trained ImageNet weights and CNN which starts with random weights. Most of the experiments were carried out on a Google Colaboratory Server using Tesla K80 GPU and Deep Graph Library v0.6.0 was used for the implementation of GCN based algorithm \cite{22}. Evaluation metrics of this study include accuracy, precision, recall and f1 score which are defined below.

\begin{figure*}[hb]
  \centering
  \subfigure[Training loss for different class classification]{
  \begin{minipage}{.45\textwidth}
  \centering
  \includegraphics[width=7.5cm]{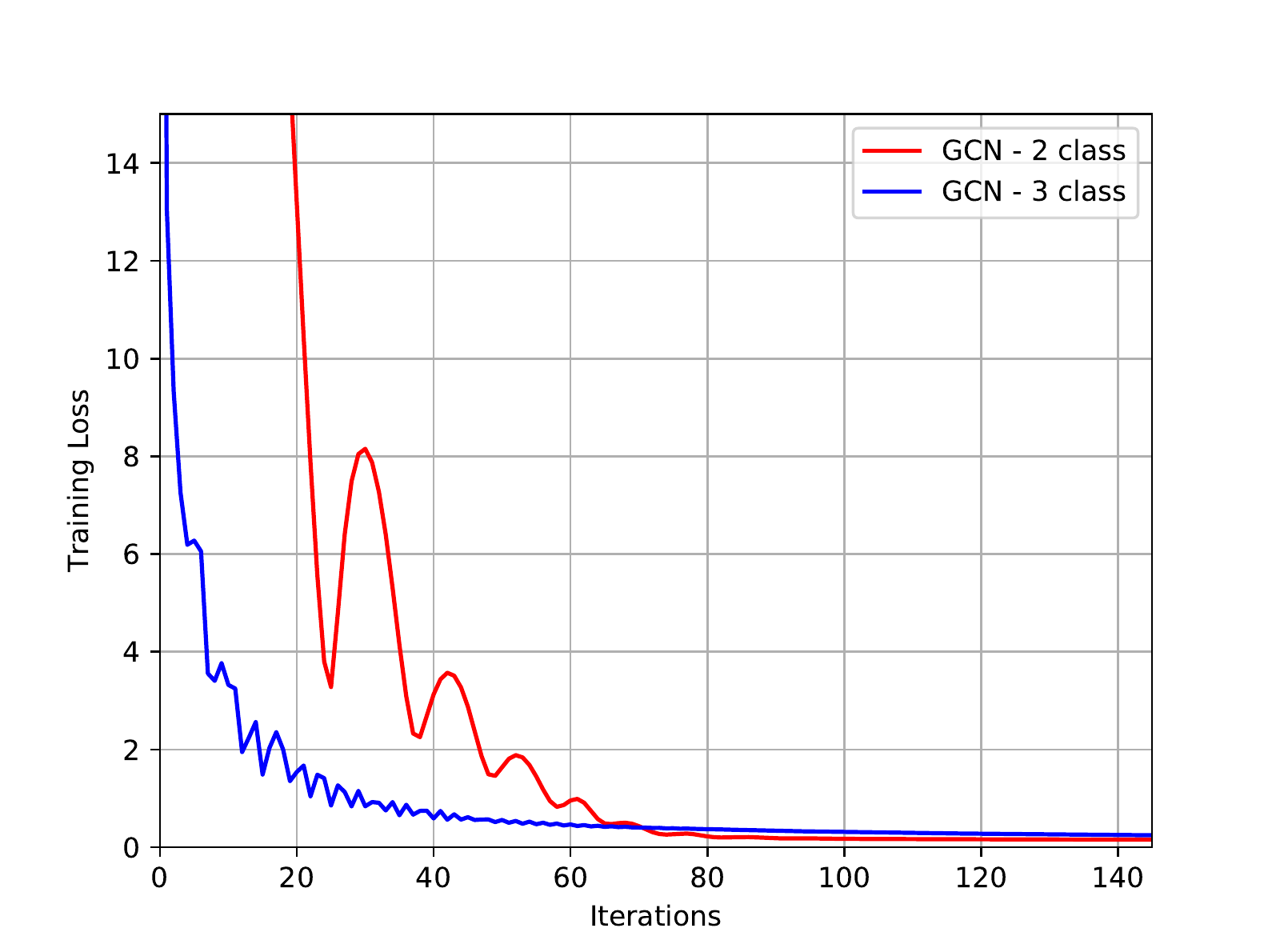}
  \end{minipage}\hfill}
  \subfigure[Testing accuracy for different class classification]{
  \begin{minipage}{.45\textwidth}
  \centering
  \includegraphics[width=7.5cm]{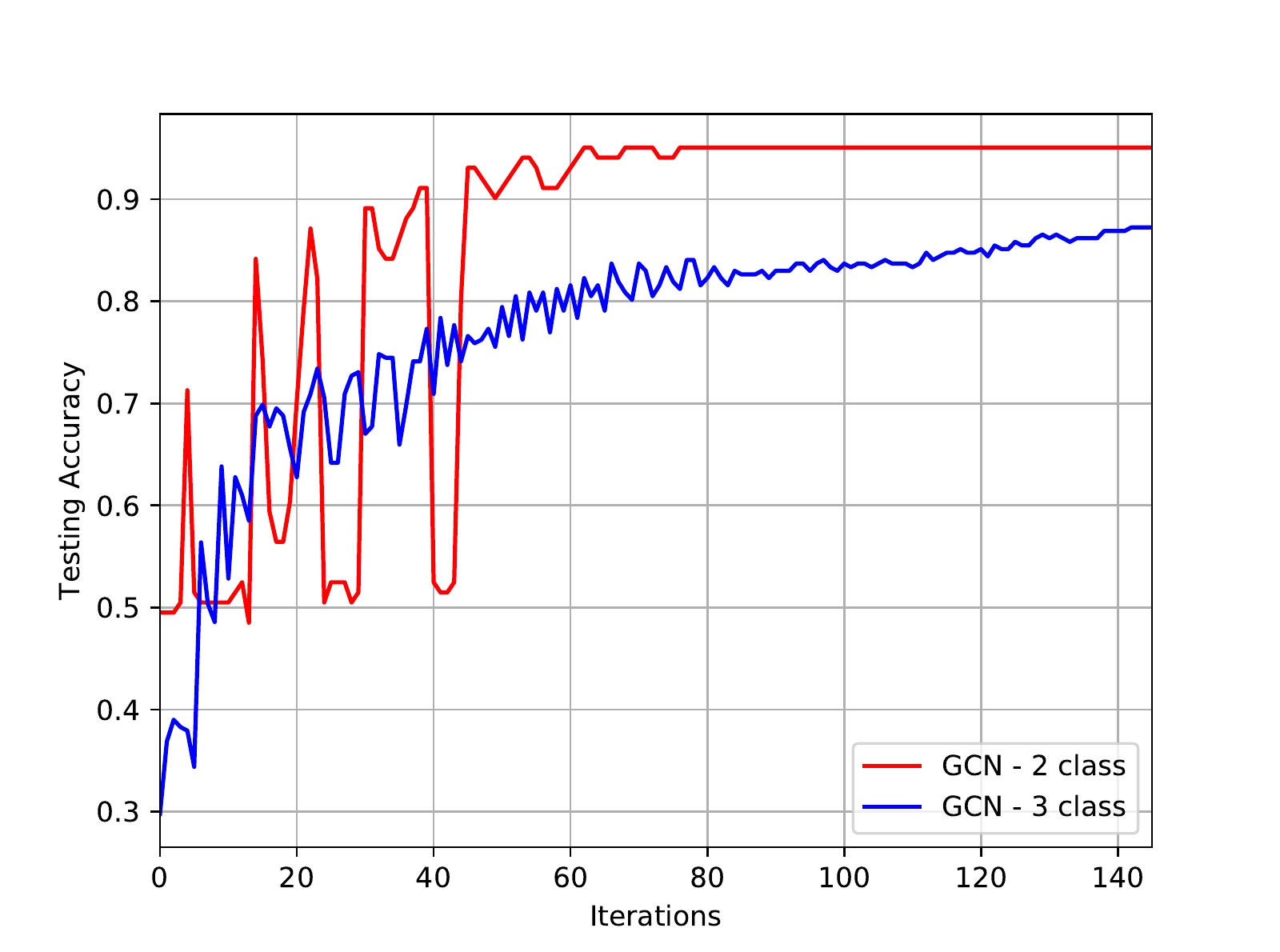}
  \end{minipage}\hfill}
  \caption{Training loss and testing accuracies for GCN}
\end{figure*}

\begin{gather*}
Accuracy=  \frac{TP+TN}{TP+TN+FP+FN}\\
Precision=  \frac{TP}{TP+FP}\\ 
Recall=  \frac{TP}{TP+FN}\\
F1score=  \frac{2TP}{2TP+FP+FN}
\end{gather*}

Where $TP$, $TN$, $FP$ and $FN$ are true positives, true negatives, false positives and false negatives respectively. 

\subsection{The training loss and testing accuracy of GCN}
\noindent

As mentioned in early sections, we created two classification problems: two class classification to identify covid CXR images from normal CXR and three class classification to differentiate covid CXR from normal and other type of Pneumonia CXR images. First we adopted a GCN with one hidden layer ($10\rightarrow20 \rightarrow3$) which has the initial feature vector dimension as 10, hidden feature vector size as 20 and output vector for three class classification. Next we compare an another GCN architecture with two hidden layers ($10\rightarrow50 \rightarrow20\rightarrow3$) for the same problem and we could observe stable performance with two hidden layer architecture considering the neighbourhood of three hops. Therefore, we adopted two hidden layer architecture for the proposed GCN model on two class and three class classifications. 

Figure 6 plots training loss and testing accuracies across 150 iterations on the above proposed GCN model. Here we can observe loss is gradually decreasing in both 2 class and 3 class problems and they are converging after $80^{th}$ iteration. Though there are fluctuations in 2 class at early iterations, 2 class hits high accuracy than 3 class due to relative simplicity of the classification task. Therefore, 2 class GCN hits about 95\% accuracy at  $80^{th}$ iteration and retrains this value after that. But GCN based 3 class problem is reaching about 83\% accuracy and slowly increasing the accuracy which implies it needs more iterations due to the hardness of differentiating Covid-19 CXR features from other pneumonia features. 

\subsection{Comparison of GCN with other models}
\noindent

In this section we compare proposed GCN method with other conventional methods which utilize transfer learnt benchmark models: ResNet-50, InceptionV3 and VGG16 on two class and three class problems separately. This performance comparison is depicted in Figure 7.  

\begin{figure*}[ht]
  \centering
  \subfigure[Training loss with epochs]{
  \begin{minipage}{.32\textwidth}
  \centering
  \includegraphics[width=6cm]{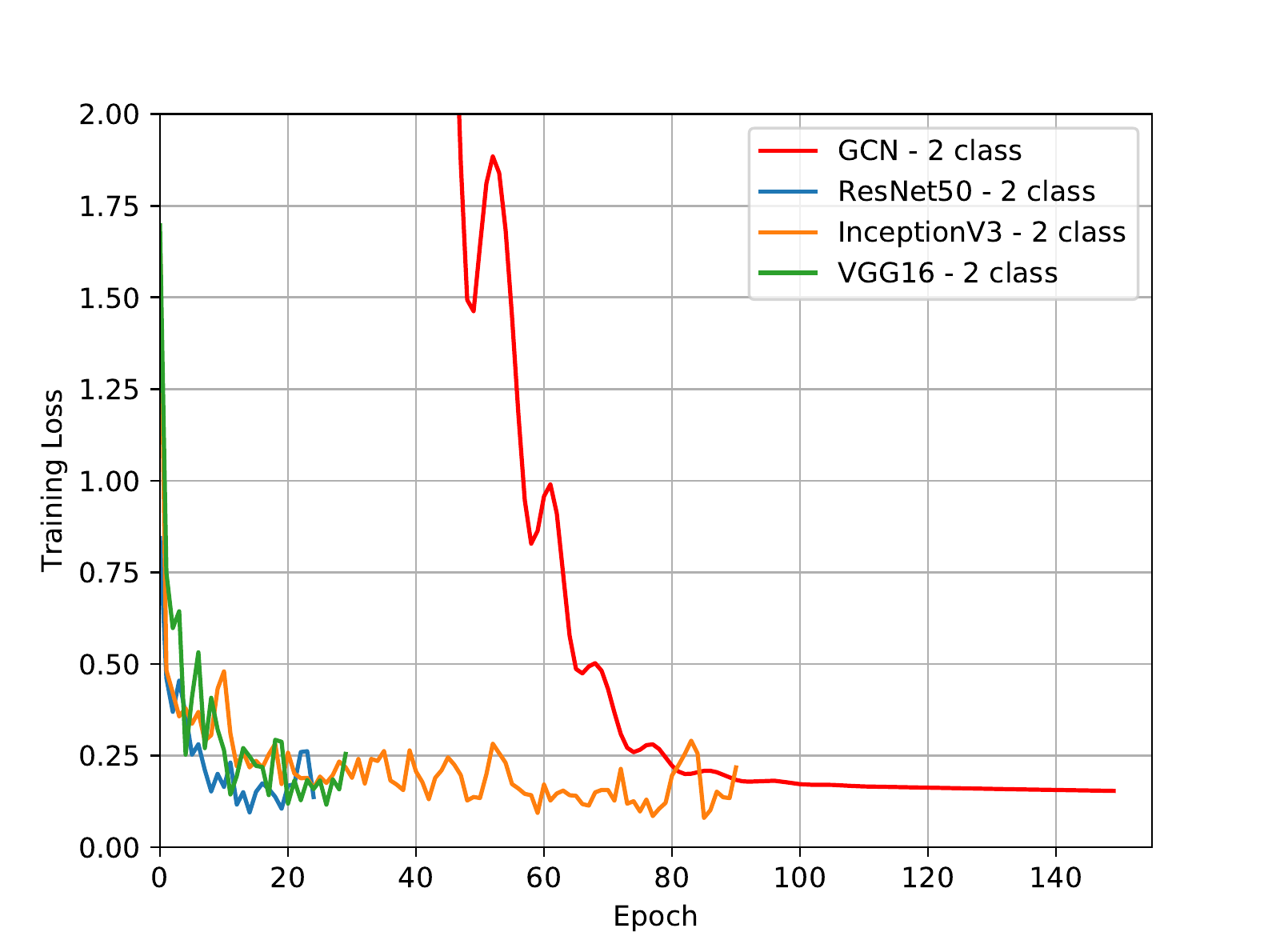}
  \end{minipage}\hfill}
  \subfigure[Training loss with time (s)]{
  \begin{minipage}{.32\textwidth}
  \centering
  \includegraphics[width=6cm]{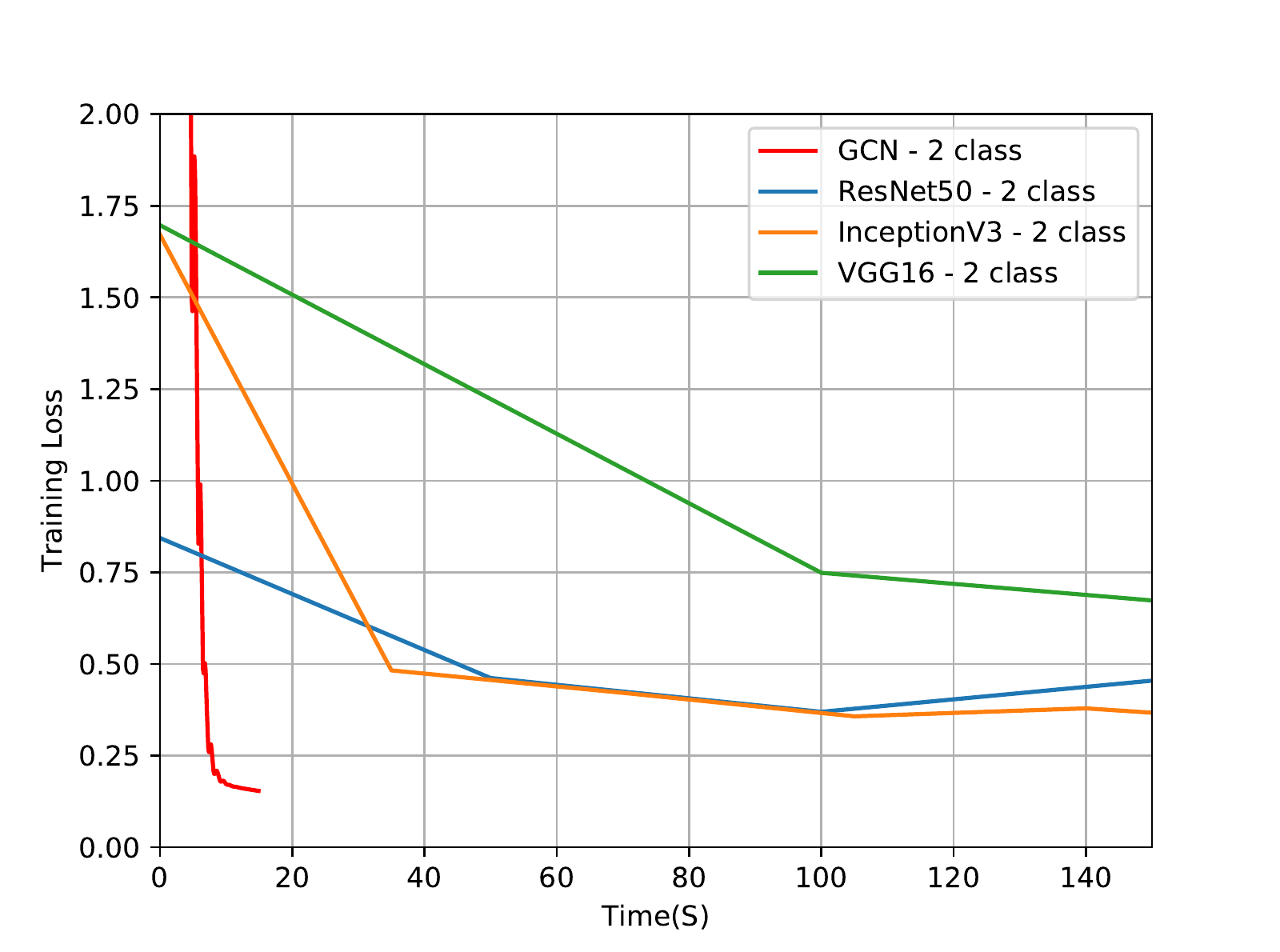}
  \end{minipage}\hfill}
  \subfigure[Testing accuracy with epochs]{
  \begin{minipage}{.32\textwidth}
  \centering
  \includegraphics[width=6cm]{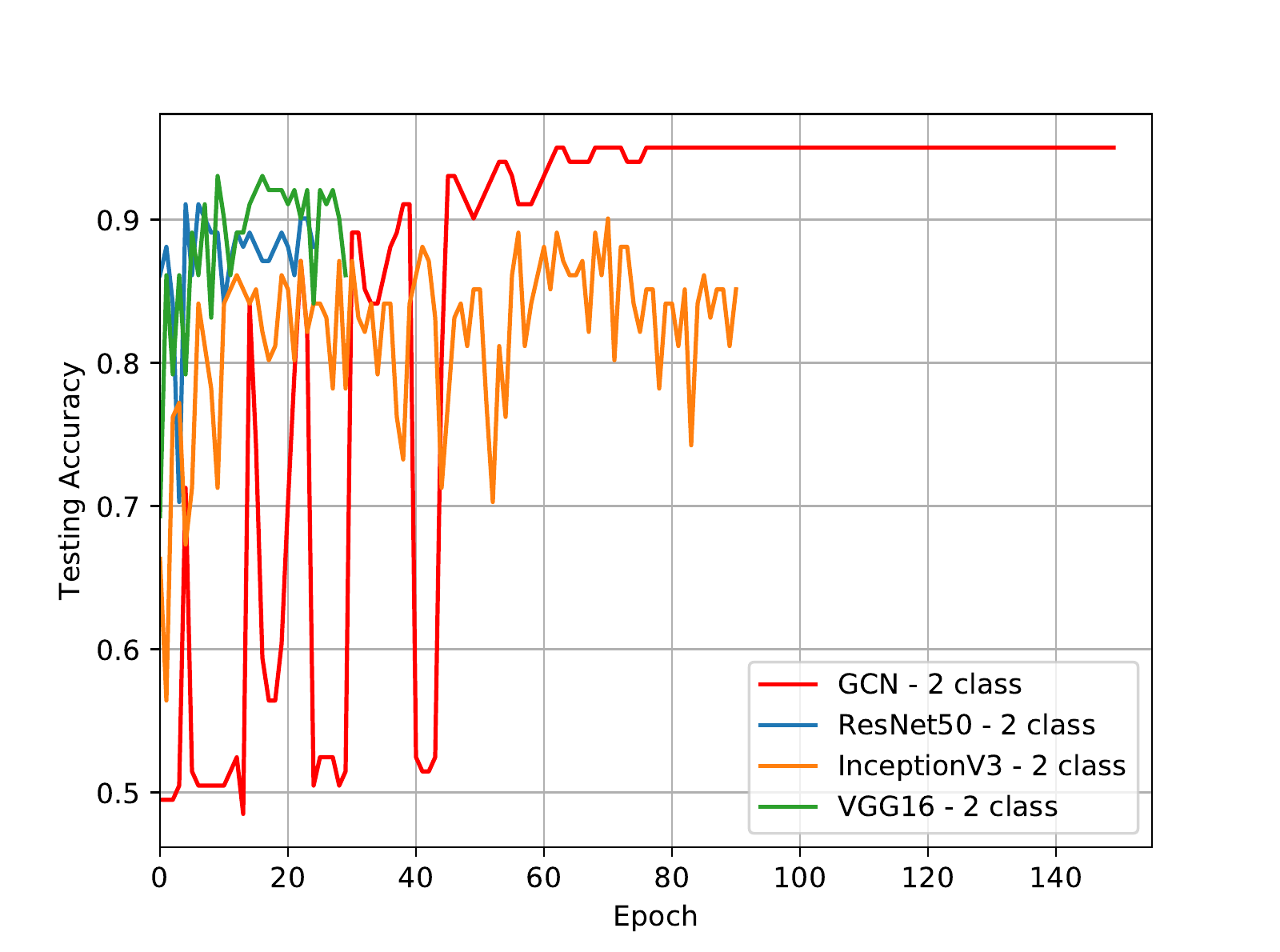}
  \end{minipage}\hfill}
  \caption{Comparison of GCN and other models on binary problem}
\end{figure*}

\begin{figure*}[ht]
  \centering
  \subfigure[Training loss with epochs]{
  \begin{minipage}{.32\textwidth}
  \centering
  \includegraphics[width=6cm]{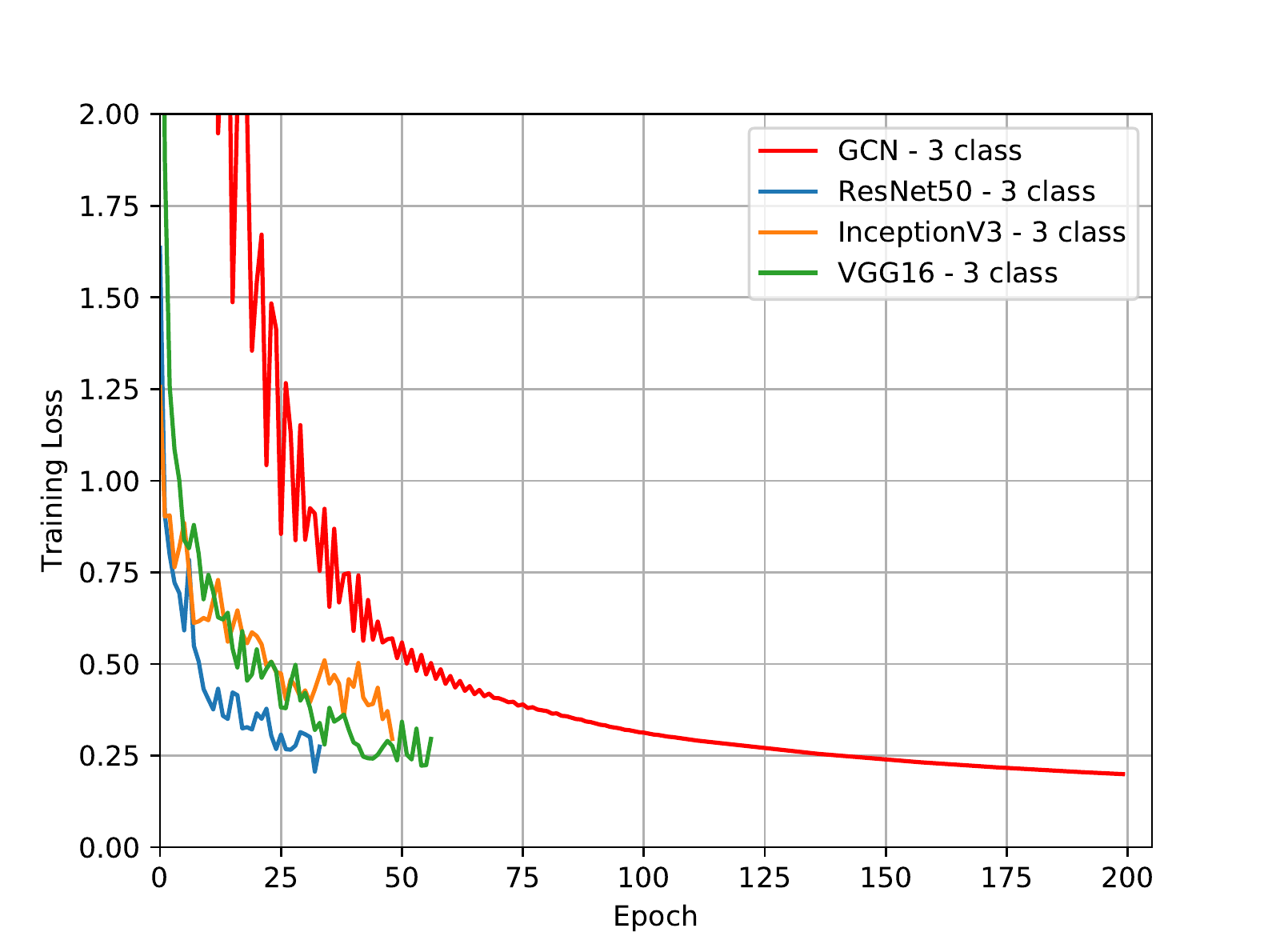}
  \end{minipage}\hfill}
  \subfigure[Training loss with time (s)]{
  \begin{minipage}{.32\textwidth}
  \centering
  \includegraphics[width=6cm]{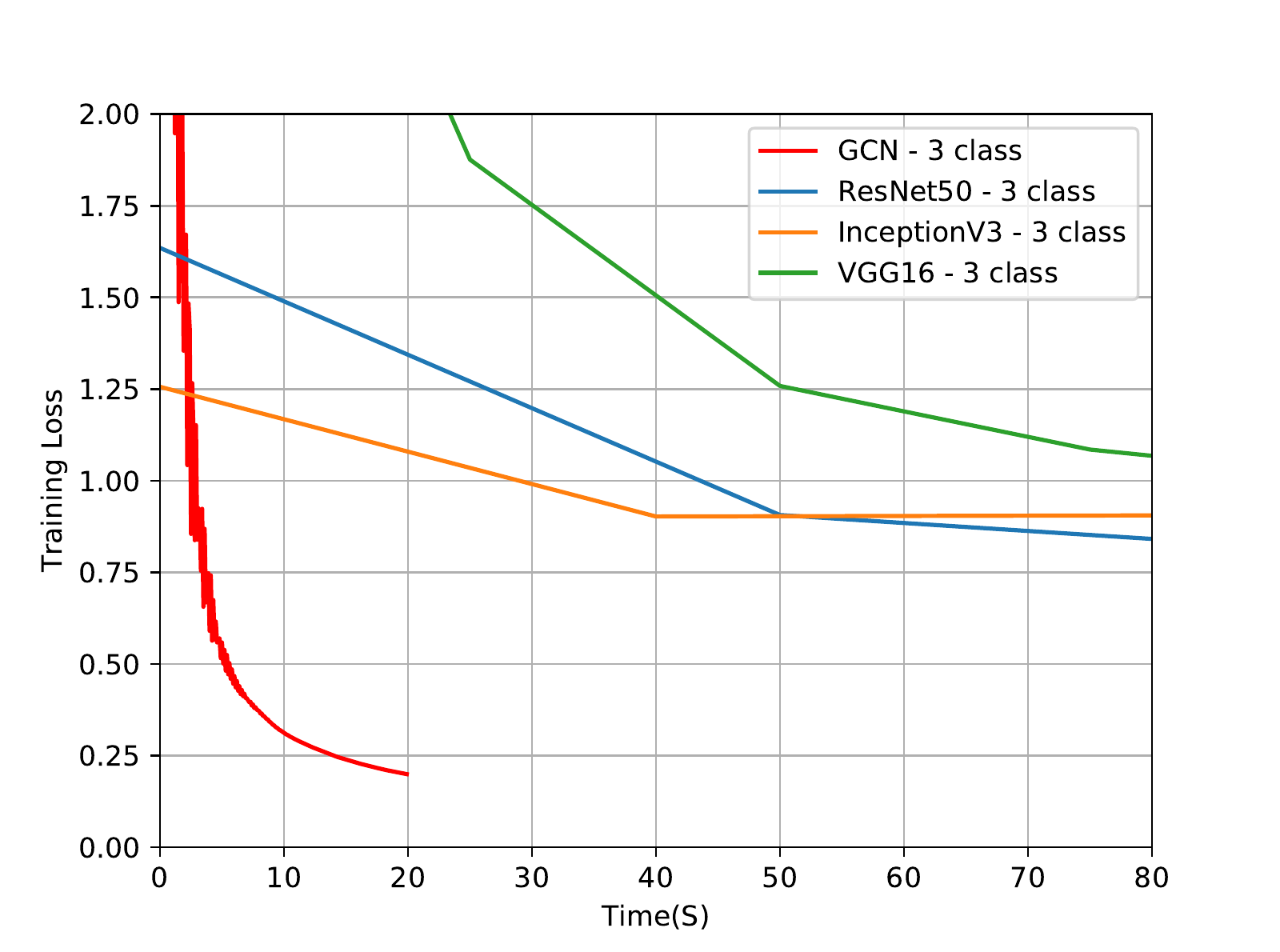}
  \end{minipage}\hfill}
  \subfigure[Testing accuracy with epochs]{
  \begin{minipage}{.32\textwidth}
  \centering
  \includegraphics[width=6cm]{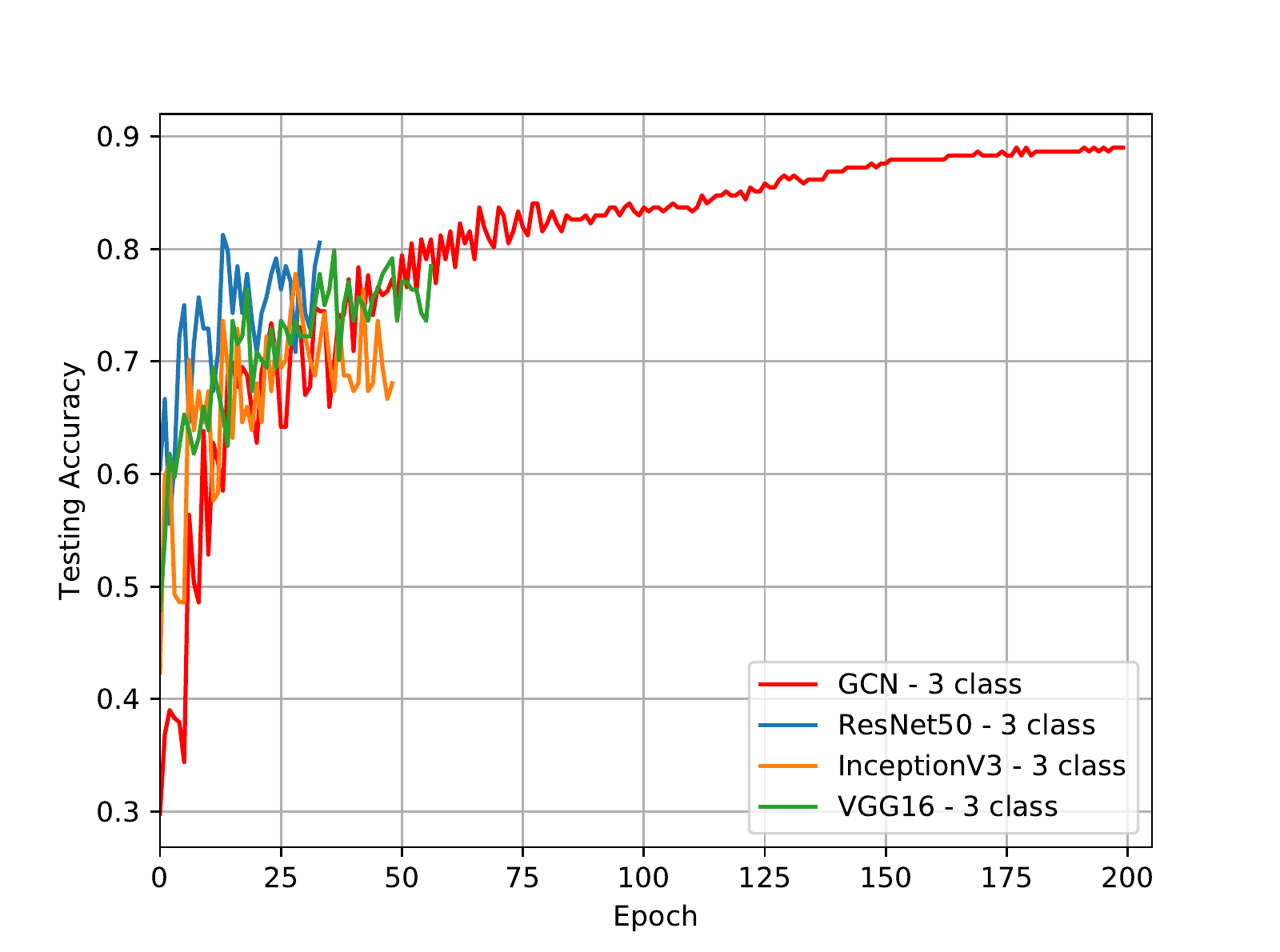}
  \end{minipage}\hfill}
  \caption{Comparison of GCN and other models on three class problem}
\end{figure*}

Here we adopted early stopping with 20 patience on benchmark models and allowed GCN model to run for about 150 iterations. Though we plot the training loss of GCN model and other benchmark models in the same plot for the comparison in Figure 7(a) and Figure 8(a) for binary and three class respectively, one iteration of GCN is not equal to one epoch in CNN models. Therefore we also show the learning curves of all the models against time axis (in seconds) as well in Figure 7(b) and Figure 8(b). In binary problem, all the transfer learnt models start with lower loss and achieve the lowest loss around 0.2 as their best performance. Meanwhile, GCN model begins with relatively higher loss but drastically reducing to its best value and converging. Specially in three class classification, GCN attains a lower loss than 0.25 while other models record their lowest loss in the range 0.25 and 0.5. GCN takes about 80 and 140 iterations to reach the 0.25 loss for binary (Figure 7(a)) and for three class (Figure 8(a))  which consume only about 15s and 20s according to Figure 7(b) and Figure 8(b). Approximately, one iteration of GCN takes less than 0.1s while one epoch of CNN model takes about 50s. Moreover, we can observe that GCN model is gradually decreasing loss in both problems and converging at later iterations while others are having fluctuations in their learning curves. We see the similar behavior in testing accuracies of both problems in GCN compared to other conventional methods in Figure 7(c) and 8(c). Exact accuracy values for GCN and other models are given in the Table 1 and Table 2 along with other metrics as well.

\begin{table}[h]
  \label{sample-table}
  \caption{Results of binary classification covid vs. Normal}
     \vskip 4mm
  \centering
  \begin{tabular}{lllll}
    \toprule
Model & ResNet-50 & InceptionV3 & VGG-16 & GCN \\
\midrule
Accuracy & 0.9108 & 0.6336 & 0.9306 & \textbf{0.9445}\\
Precision & 0.9109 & 0.7897 & \textbf{0.9324} & 0.9189\\
Recall & 0.9109 & 0.6300 & 0.9303 & \textbf{0.9764} \\
F1-score & 0.9108 & 0.5732 & 0.9305 & \textbf{0.9467}\\
    \bottomrule
  \end{tabular}
\end{table}

In Table 1, ResNet-50, VGG16 and GCN hit more than 90\% for all metrics while InceptionV3 has comparatively very low performance. Overall GCN gives the highest accuracy and highest F1-score around 94\%. Highest recall 97\% of GCN indicates fewer false negatives which is important in Covid-19 detection. Otherwise, not identifying the positive cases correctly can lead to sever consequences such as not being able to quickly isolate and treat patients which is highly required for controlling the spread. The highest precision 93\% is recorded from VGG16 while GCN is showing still better precision 91\%. However, the highest F1-score is given by GCN which is more robust measure as it encounters both precision and recall. 

\begin{table}[h]
  \label{sample-table}
  \caption{Results of 3 class classification covid, Normal and Pneumonia}
     \vskip 4mm
  \centering
  \begin{tabular}{lllll}
    \toprule
Model & ResNet-50 & InceptionV3 & VGG-16 & GCN \\
\midrule
Accuracy & 0.8125 & 0.7777 & 0.7708 & \textbf{0.8596}\\
Precision & 0.8235 & 0.7900 & 0.7759 & \textbf{0.8601}\\
Recall & 0.8078 & 0.7818 & 0.7619 & \textbf{0.8578}\\
F1-score & 0.8118 & 0.7791 & 0.7604 & \textbf{0.8586}\\
    \bottomrule
  \end{tabular}
\end{table}

In three class classification, both InceptionV3 and VGG16  have degraded performance significantly in classifying CXR into: Covid-19, normal and other pneumonia types. Only ResNet-50 and GCN achive comparatively higher performance where the highest 86\% accuracy is given by GCN. This indicates the difficulty of differentiating Covid-19 infected CXR images from other pneumonia infections for deep learning based models where relatively better performance is reported from the GCN based approach.

Next, we are exploring the accuracy, precision , recall and f1-score performance on individual classes related to binary and three class classification problems. Table 3 consists of above values for different models in binary classification problem. 

\begin{table}[h]
    \caption{Results of binary classification on individual classes}
   \vskip 4mm
  \label{sample-table}
  \centering
  \begin{tabular}{llllll}
    \toprule
\multicolumn{2}{c} {Model} & Resnet & Inceptn & VGG & GCN \\
\midrule
\multirow{4}{*}{Covid} & Acc. & 0.8613 & 0.7029 & 0.9306 & \textbf{0.9445}\\\
& Prec. & 0.9111 & 0.6363 & 0.9074 & \textbf{0.9189}\\
& Rec. & 0.8039 & 0.9607 & 0.9607 & \textbf{0.9764}\\
& F1 & 0.8541 & 0.7656 & 0.9333 & \textbf{0.9467}\\
\midrule
\multirow{4}{*}{Normal} & Acc. & 0.8613 & 0.7029 & 0.9306 & \textbf{0.9445}\\\
& Prec. & 0.8214 & 0.9166 & 0.9574 & \textbf{0.9747}\\
& Rec. & \textbf{0.9200} & 0.4400 & 0.9000 & 0.9120\\
& F1 & 0.8679 & 0.5945 & 0.9278 & \textbf{0.9421}\\
    \bottomrule
  \end{tabular}
\end{table}

We observe the highest performance for both classes covid and normal is given by GCN model where only the recall for normal class of ResNet-50 is marginally higher than GCN. But ResNet-50 has poor performance for most of the other metrics around 80\% - 86\% in both classes while GCN is showing relatively higher and stable around 94\% to 97\% for most of the metrics.  

\begin{table}[h]
    \caption{Results of three class classification on individual classes}
   \vskip 4mm
  \label{sample-table}
  \centering
  \begin{tabular}{llllll}
    \toprule
\multicolumn{2}{c} {Model} & Resnet & Inceptn & VGG & GCN \\
\midrule
\multirow{4}{*}{Covid} & Acc. & 0.8333 & 0.8541 & 0.8402 & \textbf{0.8803}\\
& Prec. & 0.7288 & \textbf{0.8260} & 0.7692 & 0.8238\\
& Rec. & 0.8431 & 0.7450 & 0.7843 & \textbf{0.8478}\\
& F1 & 0.7818 & 0.7835 & 0.7766 & \textbf{0.8346}\\
\midrule
\multirow{4}{*}{Normal} & Acc. & 0.9236 & 0.8819 & 0.8750 & \textbf{0.9496}\\\
& Prec. & \textbf{0.9534} & 0.9459 & 0.7666 & 0.9523\\
& Rec. & 0.8200 & 0.7000 & \textbf{0.9200} & 0.8991\\
& F1 & 0.8817 & 0.8045 & 0.8363 & \textbf{0.9248}\\
\midrule
\multirow{4}{*}{Pneumonia} & Acc. & 0.8750 & 0.8402 & 0.8333 & \textbf{0.8801}\\\
& Prec. & \textbf{0.8378} & 0.6923 & 0.8064 & 0.7926\\
& Rec. & 0.7209 & \textbf{0.8372} & 0.5813 & 0.8112\\
& F1 & 0.7750 & 0.7578 & 0.6756 & \textbf{0.8007}\\
    \bottomrule
  \end{tabular}
\end{table}

Table 4 reports the individual class performance for all transfer learnt model and GCN on three class classification. Relatively low performance in covid and pneumonia classes compared to normal class is visible related to all the models in Table 4 which indicates the difficulty of differentiating covid infected CXR from the patterns seen on CXR images of other infections. But still most of the highest and better performance is given by GCN based approach where few deviations to this can be noted with only small differences. Specially, for all the classes covid, pneumonia and normal, the highest accuracy and F1-score is from GCN. Another important observation is VGG16 and ResNet-50 which are relatively good performing models in binary classification has significant performance degradation when the difficulty of classification is raised from binary to three class. However, GCN maintains the highest performance in most metrics for both problems.

\subsection{Different aggregation functions and connectivity thresholds}
\noindent

There are two main steps in GCN algorithm called: "Aggregation" and "Update". Meanwhile, $\alpha$ is an important threshold which decides the connectivity of two nodes based on the encoded feature similarity values. The purpose of this experiment is to measure the GCN based model performance with different aggregation functions across a range of $\alpha$ values. The results of above study on binary classification problem is depicted in Figure 9.

\begin{figure}[h]
  \centering
  \includegraphics[width=0.45\textwidth]{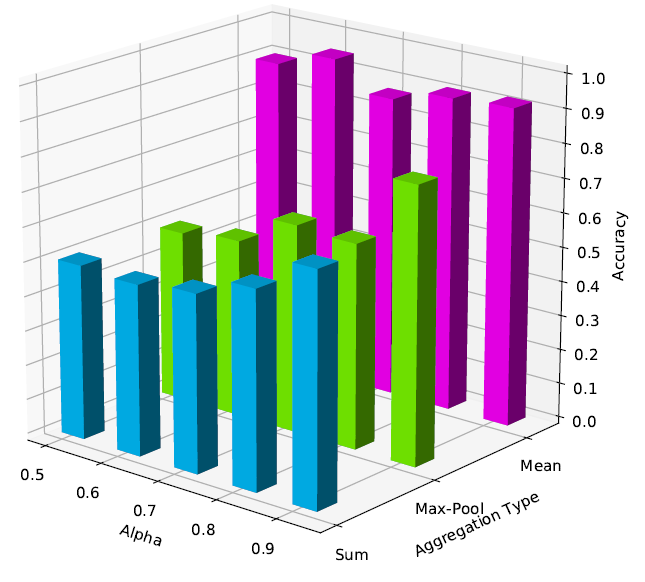}
  \caption{Accuracy for different aggregation functions and different $\alpha$ values for binary classification}
\end{figure}

It is clearly visible that mean aggregation function has the highest accuracy closer to 90\% or better for all $\alpha$ values while summation and max pooling give poor performance. There is little incremental improvement for max pooling and summation aggregations while the graph density is decreasing. Still the accuracies are very low, around 60\% to 70\% for the graph with the lowest density ($\alpha = 0.9$). This experiment on three class classification was also performed and is shown in the Figure 10. 

\begin{figure}[h]
  \centering
  \includegraphics[width=0.45\textwidth]{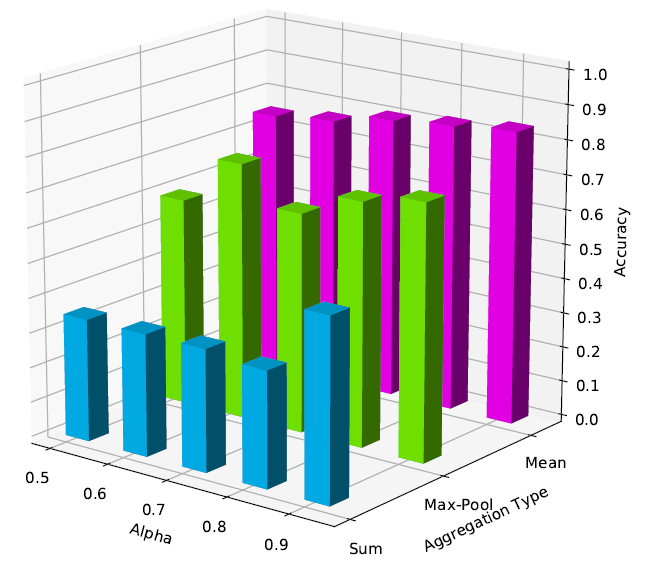}
  \caption{Accuracy for different aggregation functions and different $\alpha$ values for three class classification }
\end{figure}

In three class classification, summation aggregation shows poor performance while max pooling is relatively better than summation. However, the highest accuracy for the range of thresholds $\alpha$ is again corresponding to mean aggregation. Though we observe little improvement while the density of the graph is reducing in Figure 10, the highest accuracy can be achieved for a graph with higher density ($\alpha = 0.6$) in Figure 9 for the binary problem. This proves the generalization capability of mean aggregation function independent of the irregularities of graph data like variable and unordered neighbours for any given node in the graph. Further, the mean aggregation is demonstrating its ability to learn highly densed graphs with more edges in a simple calcification like binary (Figure 9) where as for in three class classification the performance can be affected to some extent due to the hardness of the classification problem (Figure 10).

\subsection{Five class classification performance and saliency map for encoder}
\noindent

\begin{figure*}[ht]
  \centering
  \subfigure[ResNet-50 output]{
  \begin{minipage}{.45\textwidth}
  \centering
  \includegraphics[width=7.5cm]{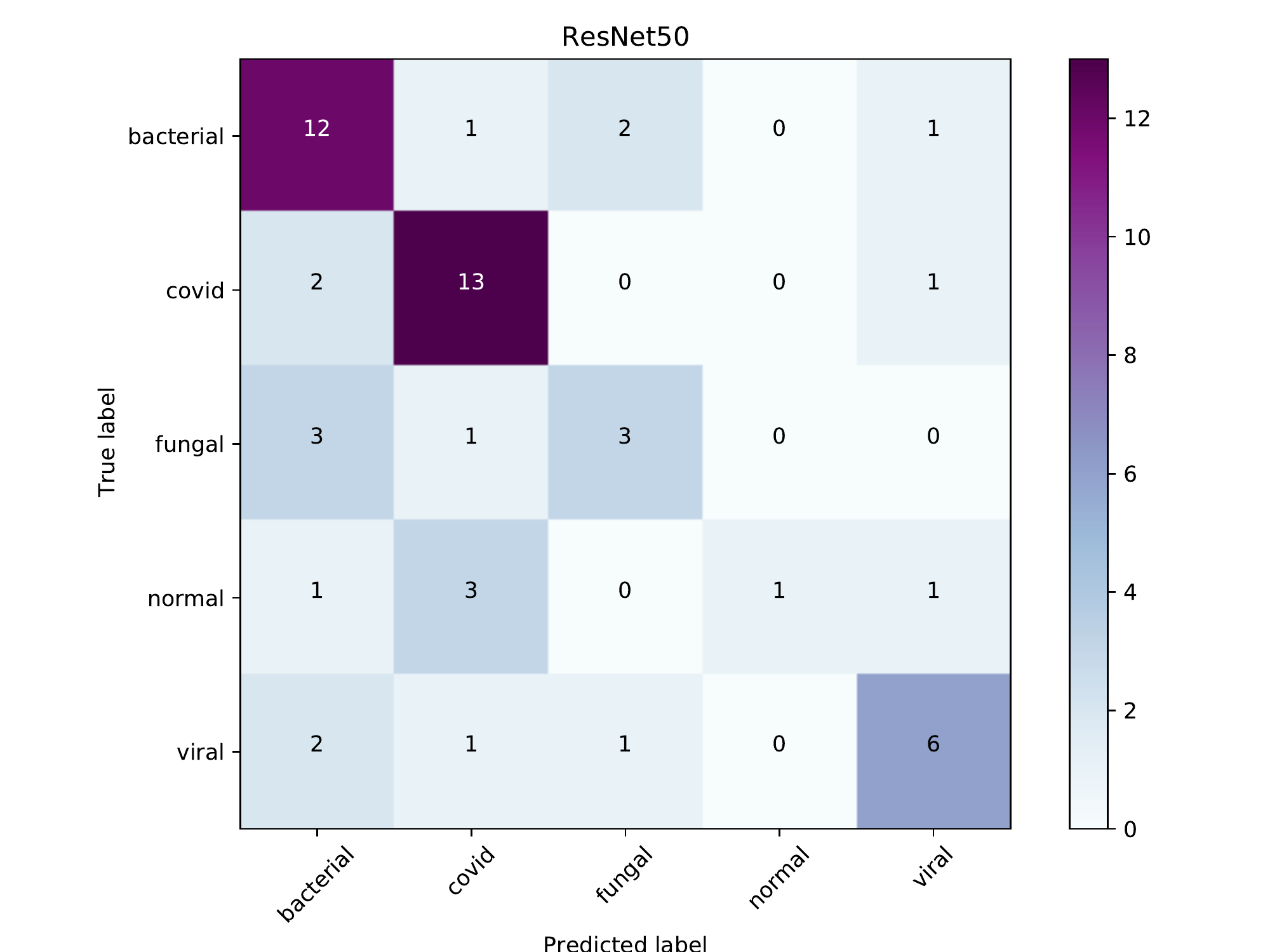}
  \end{minipage}\hfill}
  \subfigure[GCN output]{
  \begin{minipage}{.45\textwidth}
  \centering
  \includegraphics[width=7.5cm]{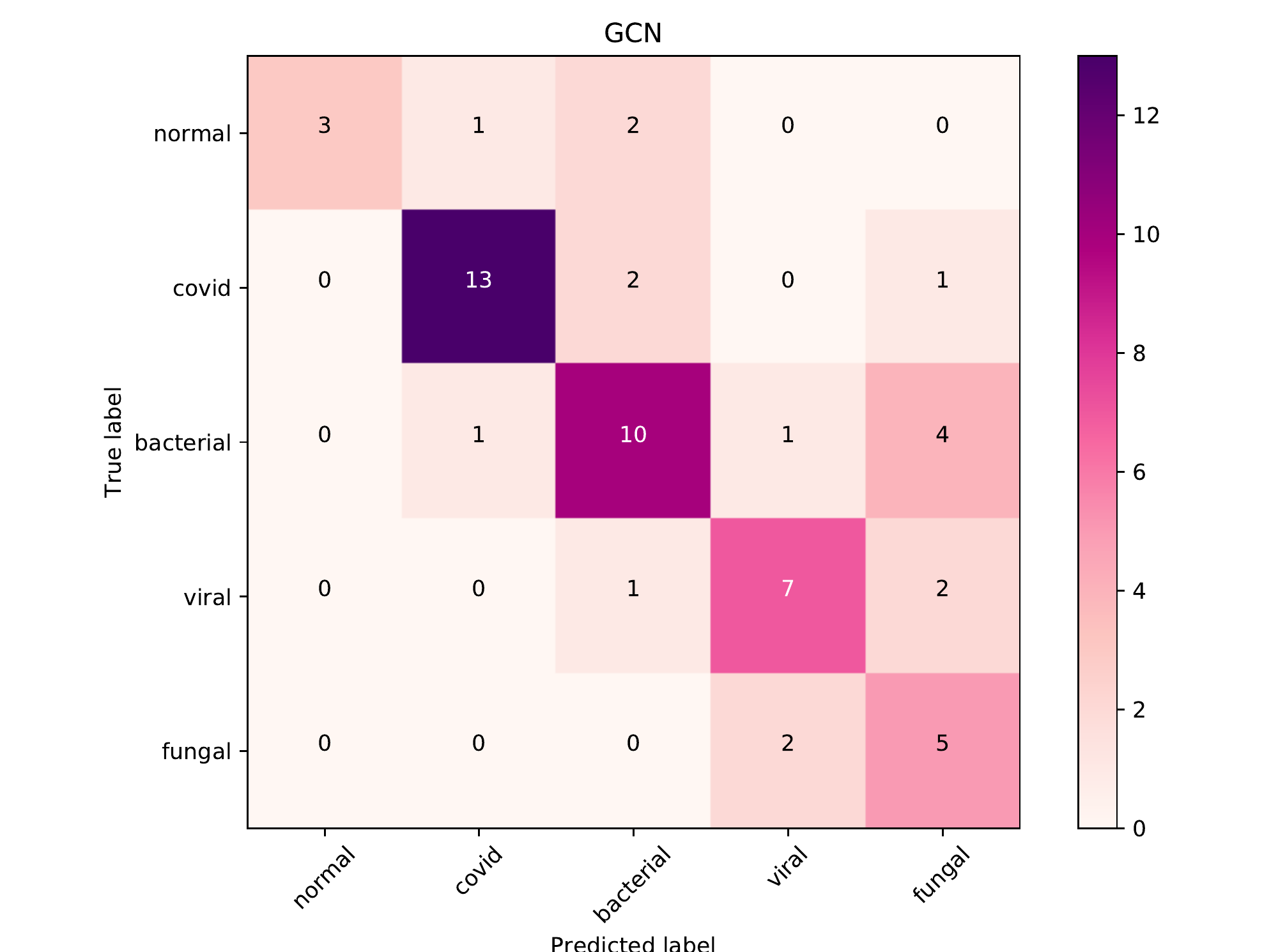}
  \end{minipage}\hfill}
  \caption{Confusion matrices for five class classification}
\end{figure*}

We further evaluated the performance of the proposed GCN based model on a harder classification with five classes: Covid-19, bacterial, fungal, other viral and normal. The purpose of this experiment is to measure the capabilities of GCN model in differentiating covid-19 infection patterns from other types of bacterial, fungal and other viral types' infection patterns. Here we compare the outcome of GCN model on above five class classification against transfer learnt ResNet-50 on the same subjects using confusion matrices and given in Figure 11. 

When we compare confusion matrices of Figure 11 (a) and (b), the highest number of true positives for all classes is given by GCN where ResNet-50 is having low true positives. GCN has predicted more than or equal positive cases correctly compared to ResNet-50 for all classes except bacterial. Most importantly, we notice a distribution of zero predictions under the predicted label covid, bacterial and viral in Figure 11 (b) which indicates less false positives where as ResNet-50 is having higher number of false positive predictions such as viral, normal, fungal and bacterial CXR as covid infected. We see the same behaviour for bacterial class as well in Figure 11(a). Thus we can conclude the proposed GCN based model with less training time is showing better performance compared to ResNet-50 in this five class classification. 

\begin{figure*}[h]
  \centering
  \subfigure{
  \begin{minipage}{.95\textwidth}
  \centering
  \includegraphics[width=15cm]{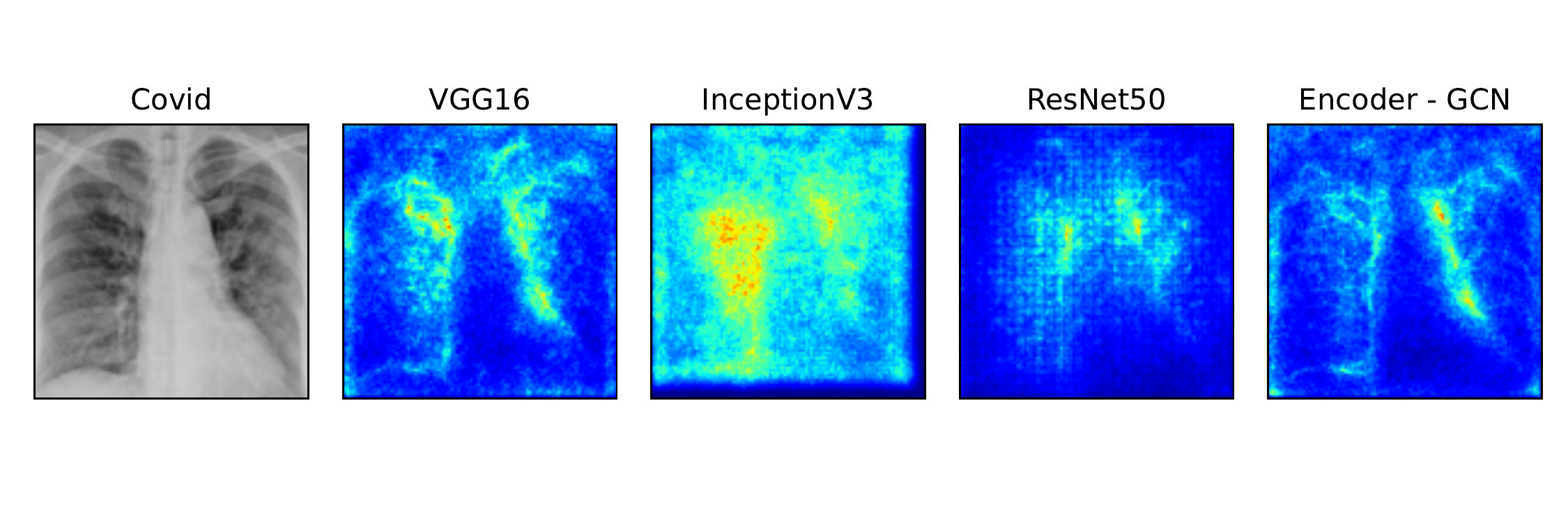}
  \end{minipage}\hfill}
  \caption{Sailency maps related to VGG16, InceptionV3, ResNet-50 and CNN encoder for GCN corresponding to a given Covid CXR}
\end{figure*}

Next, we did another experiment to observe the efficiency of the CNN encoder that we used to extract the features of CXR images and which we we then feed to build the graph. For that, we drew saliency maps which is an important concept of deep learning where the spatial distribution of saliency towards the final classification is proportional to the brightness of pixels in the map. Here we compare the proposed CNN encoder against other transfer learnt model VGG16, ResNet-50 and InceptionV3 using saliency maps and they are given in Figure 12. 

In Figure 12, the most important regions which were utilized for final prediction is highlighted related to each model. According to them, InceptionV3 has not focused on important regions and the prediction is guided by pixels all over the image. On the other hand, ResNet-50 is not correctly guided by all areas with hazzy opacification according to the highlighted attended locations. The encoder used in GCN is better than VGG16 as its classification is more focused and guided by the areas with cloudy visibility in the given covid CXR image whereas VGG16 is showing more scattered attention. 

\section{Conclusion}
\noindent

Covid-19 is a major health problem that has added more challenges to people's lives which appeals for innovative and efficient approaches for tackling these kind of problems. One instance is that the demand for effective automated disease diagnosis methods such as medical image processing as a solution for resources limitations of tests like RT-PCR. One of the frequently used methods for medical image processing is transfer learning where the performance of these models are affected by various constraints. Also, existing methods are limited for binary classification while there is an appealing need for models to differentiate Covid-19 from other lung infections. As a solution, a GCN based novel method is proposed in this work to classify CXR images into binary and three class classifications. The strength of the proposed model leverages on the combination of graph representation and convolution operation. Thus, it exploits not only the data but also important relational knowledge between data instances and also utilizes metadata information as features in the proposed model. A CNN encoder is used to convert CXR into feature vectors which are then used to build the graph based on the similarity matrix between data instances. GCN algorithm is adopted on the built graph where the final high level features are classified into different classes using a softmax layer. Multiple experiments on two class and three class classifications of proposed GCN model show better performance compared to conventional transfer learnt models ResNet-50, InceptionV3 and VGG16. More specifically GCN reports the highest accuracy 94\% and 86\% in binary and three class predictions respectively. Further we observe the mean aggregation function has more generalization capabilities hence better performance in graph data learning other than sum and max pool aggregation even in a more densed graph. Moreover, the classification performance on a five class problem has been compared to a ResNet-50 model on the same subjects and GCN is showing high true positives and less false positives using confusion matrices. This is an important fact as having correctly identifying positive cases as much as possible is critical for controlling the spread of a disease. The correctness of extracted features which we feed for the graph is evaluated using saliency maps next. Identifying infected areas of given CXR images by proposed CNN encoder further highlights the effectiveness of proposed novel GCN model. 


\textbf{References}}

\begin{thebibliography}{99}
\zihao{5-} \addtolength{\itemsep}{-1em}
\vspace {1.5mm}
 

\bibitem[1]{1}
J. Deng, W. Dong, R. Socher, L.-J. Li, K. Li, and L. Fei-Fei, “Imagenet: A large-scale hierarchical image database,” in 2009 IEEE conference on computer vision and pattern recognition. Ieee, 2009, pp. 248–255.

\bibitem[2]{2}
I. D. Apostolopoulos and T. A. Mpesiana, “Covid-19: automatic detection from x-ray images utilizing transfer learning with convolutional neural networks,” Physical and Engineering Sciences in Medicine, vol. 43, no. 2, pp. 635–640, 2020.

\bibitem[3]{3}
N. N. Das, N. Kumar, M. Kaur, V. Kumar, and D. Singh, “Automated deep transfer learningbased approach for detection of covid-19 infection in chest x-rays,” Irbm, 2020.

\bibitem[4]{4}
A. Narin, C. Kaya, and Z. Pamuk, “Automatic detection of coronavirus disease (covid-19) using x-ray images and deep convolutional neural networks,” Pattern Analysis and Applications, pp. 1–14, 2021.

\bibitem[5]{5}
A. Jaiswal, N. Gianchandani, D. Singh, V. Kumar, and M. Kaur, “Classification of the covid- 19 infected patients using densenet201 based
deep transfer learning,” Journal of Biomolecular Structure and Dynamics, pp. 1–8, 2020.

\bibitem[6]{6}
X. Zhang, S. Lu, S.-H. Wang, X. Yu, S.-J. Wang, L. Yao, Y. Pan, and Y.-D. Zhang, “Diagnosis of covid-19 pneumonia via a novel deep learning
architecture,” Journal of Computer Science and Technology, p. 1, 2021.

\bibitem[7]{7}
T. D. Pham, “Classification of covid-19 chest xrays with deep learning: new models or fine tuning?” Health Information Science and Systems, vol. 9, no. 1, pp. 1–11, 2021.

\bibitem[8]{8}
M. Nishio, S. Noguchi, H. Matsuo, and T. Murakami, “Automatic classification between covid-19 pneumonia, non-covid-19 pneumonia, and the healthy on chest x-ray image: combination of data augmentation methods,” Scientific reports, vol. 10, no. 1, pp. 1–6, 2020.

\bibitem[9]{9}
C. Sitaula and M. B. Hossain, “Attention-based vgg-16 model for covid-19 chest x-ray image classification,” Applied Intelligence, pp. 1–14, 2020.

\bibitem[10]{10}
Y. Oh, S. Park, and J. C. Ye, “Deep learning covid-19 features on cxr using limited training data sets,” IEEE Transactions on Medical Imaging, vol. 39, no. 8, pp. 2688–2700, 2020.

\bibitem[11]{11}
L. Wang, Z. Q. Lin, and A. Wong, “Covid-net: A tailored deep convolutional neural network design for detection of covid-19 cases from chest x-ray
images,” Scientific Reports, vol. 10, no. 1, pp. 1–12, 2020.

\bibitem[12]{12}
S. Basodi, C. Ji, H. Zhang, and Y. Pan, “Gradient amplification: An efficient way to train deep neural networks,” Big Data Mining and Analytics, vol. 3, no. 3, pp. 196–207, 2020.

\bibitem[13]{13}
X. Lei, J. Tie, and Y. Pan, “Inferring metabolitedisease association using graph convolutional networks,” IEEE/ACM Transactions on Computational Biology and Bioinformatics, 2021.

\bibitem[14]{14}
Y. Zhang, X. Lei, Y. Pan, and W. Pedrycz, “Prediction of disease-associated circrnas via circrna–disease pair graph and weighted nuclear norm minimization,” Knowledge-Based Systems, vol. 214, p. 106694, 2021.

\bibitem[15]{15}
T. B. Mudiyanselage, X. Lei, N. Senanayake, Y. Zhang, and Y. Pan, “Graph convolution networks using message passing and multisource similarity features for predicting circrna disease association,” in 2020 IEEE International Conference on Bioinformatics and Biomedicine (BIBM). IEEE, 2020, pp. 343–348.

\bibitem[16]{16}
X.-J. Lei, C. Bian, and Y. Pan, “Predicting circrna disease associations based on improved weighted biased meta-structure,” Journal of Computer
Science and Technology, vol. 36, no. 2, pp. 288–298, 2021.

\bibitem[17]{17}
Z. Wu, S. Pan, F. Chen, G. Long, C. Zhang, and S. Y. Philip, “A comprehensive survey on graph neural networks,” IEEE transactions on neural
networks and learning systems, 2020.

\bibitem[18]{18}
T. N. Kipf and M. Welling, “Semi-supervised classification with graph convolutional networks,” arXiv preprint arXiv:1609.02907, 2016.

\bibitem[19]{19}
W. L. Hamilton, R. Ying, and J. Leskovec, “Inductive representation learning on large graphs,” arXiv preprint arXiv:1706.02216, 2017.

\bibitem[20]{20}
J. P. Cohen, P. Morrison, and L. Dao, “Covid-19 image data collection,” arXiv 2003.11597, 2020. [Online]. Available: https://github.com/ieee8023/covid-chestxray-dataset

\bibitem[21]{21}
T. Rahman, “Covid-19 radiography database,” Mar 2021. [Online]. Available:
https://www.kaggle.com/tawsifurrahman/covid19-radiography-database

\bibitem[22]{22}
M. Wang, D. Zheng, Z. Ye, Q. Gan, M. Li, X. Song, J. Zhou, C. Ma, L. Yu, Y. Gai, T. Xiao, T. He, G. Karypis, J. Li, and Z. Zhang, “Deep graph library: A graph-centric, highlyperformant package for graph neural networks,” arXiv preprint arXiv:1909.01315, 2019.

 \end{thebibliography}

\renewcommand\refname{\zihao{5}

\begin{strip}
\end{strip}

\mbox{}
\clearpage
\clearpage

\end{document}